\documentclass[a4paper,11pt]{article}
\usepackage{pos}

\title{Confining strings, axions and glueballs in the planar limit}

\author{Andreas Athenodorou}

\affiliation{Computation-based Science and Technology Research Center, The Cyprus Institute, Cyprus}
\affiliation{\& Dipartimento di Fisica, Universit\'a di Pisa and INFN, Sezione di Pisa, Largo Pontecorvo 3, 56127 Pisa, Italy}

\emailAdd{a.athenodorou@cyi.ac.cy}

\abstract{We present recent results on the spectrum of a confining flux-tube that is closed around a spatial torus as a function of its length as well as the spectrum of glueballs. The extraction of the spectra has been realized by simulating four dimensional $SU(N)$ gauge theories and performing measurements using lattice techniques. Regarding flux-tubes, we have performed calculations for $N=3,5,6$ and for various values of spin, parity and longitudinal momentum. Long flux-tubes can be thought of as infinitesimally thin strings; hence their spectrum is expected to be described by an effective string theory. Furthermore, the flux-tube's internal structure makes possible the existence of massive states in addition to string modes. Our calculations demonstrate that although most states exhibit a spectrum which can be approximated adequately by Nambu-Goto there is strong evidence for the existence of a massive axion on the world-sheet of the QCD flux-tube as well as a bound state of two such axions. Regarding glueballs, we extracted spectra from $N=2$ to $N=12$ which enables us to extrapolate to $N= \infty$. Our main aim was to calculate the lightest glueball masses for all different configurations of the quantum numbers of spin, parity and charge conjugation. This provides a major update on the spectrum of glueballs in the planar limit.  
}

\FullConference{%
  Corfu Summer Institute 2021 "School and Workshops on Elementary Particle Physics and Gravity"\\
  29 August - 9 October 2021\\
  Corfu, Greece
}

\begin{document}
\maketitle

\section{Introduction}
\label{sec:introduction}

During the last three decades lattice gauge theory simulations provided useful information towards the physics of the 't Hooft's large-$N$ limit of gauge theories as well as QCD. In parallel, the "second superstring revolution" of Maldacena’s AdS/CFT correspondence bloomed leading to gauge-gravity dualities. Such dualities between weakly coupled string theories and strongly coupled gauge theories at large-$N$, have led to a common interest in what the physics of the large-$N$ gauge theories is.

Understanding the large-$N$ limit of gauge theories requires the investigation of masses of associated states. The simplest such states one can consider are gluballs and flux-tubes, with both states reflecting hadronic dynamics. The calculation of the spectrum of these excitations has been a matter of investigation by both, lattice gauge theories as well as strings including AdS/CFT duality and effective bosonic string theory. In addition a more straightforward relation between these two fields has been established: lattice provides data extracted considering first principles for comparison with strings.

In QCD the quarks are confined in bound states by forming open flux-tubes. Long flux-tubes behave similarly to thin strings: If you pull the string apart, at some point it breaks; thus the term confining strings. However, to observe such a phenomenon in a lattice QCD calculation it requires the introduction of dynamical quarks (sea quarks) in the Markov-chain simulation used in production of configurations. We consider pure gauge theories where such effects do not appear. By placing the confining flux-tube in a given position in space we expect $D-2$ massless modes to propagate along the string arising from the spontaneously broken translation invariance in the $D-2$ directions transverse to the flux. We, thus, expect that there should be a low energy effective string theory describing such oscillating modes. Although, a flux-tube can be considered effectively as a thin string, it also has an intrinsic width. This suggests that massive states related to the intrinsic structure of the tube may exist in the spectrum. One can investigate, whether, such states exist by extracting the flux-tube spectrum, compare it with an effective string theory model and identify states which exhibit significant deviations from a theoretical description. A naive expectation would be that a massive mode has the characteristics of a resonance with energy gap of the order of the mass gap (scalar glueball mass $\sim m_G$) of the theory. For reasons of simplicity, we investigated the spectrum of the closed flux-tube which winds around the spatial lattice torus, thus the name "torelon". This set up avoids the consideration of the effect of the static quarks on the spectrum, and focuses on the dynamics of the flux-tube.

In the past it has been demonstrated~\cite{Athenodorou:2011rx} that the confining string in $D=2+1$ $SU(N)$ gauge theories can be adequately approximated by the Nambu-Goto free string in flat space-time, from short to long flux-tubes, without any massive excitations showing up. Furthermore, we demonstrated~\cite{Athenodorou:2010cs} that the spectrum of the closed flux-tube in $D=3+1$ consists mostly of string-like states but in contrast to the $D=2+1$ case a number of excitations with quantum number $J^{P} = 0^-$ appeared to be in accordance with the characteristics of a massive excitation. In 2013, Dubovsky et al,~\cite{Dubovsky:2013gi} demonstrated that this state arises naturally if one includes a Polyakov topological piece in the string theoretical action. Our old results were poor - the spectrum has been extracted for a few string lengths, and for low statistics. Recently, we proceeded towards a major improvement of the previous investigation on $D=3+1$ by extracting the spectrum of the flux-tube for all the irreducible representations expanded by the quantum numbers (QNs) $\{ |J|,P_{\perp},P_{\parallel} \}$ using three values of color $N$, namely $N=3,5,6$, as well as by probing through a large set of flux-tube lengths.

In addition to flux-tubes we have also improved the older glueball spectra calculations in the Large-$N$ limit. Our main aim in this work is to provide a calculation of the low-lying ‘glueball’ mass spectrum for all quantum numbers and all values of $N$. This means calculating the lowest states in all the irreducible representions, $R$, of the rotation group of a cubic lattice, and for both values of parity $P$ and charge conjugation symmetry $C$. We do so by performing calculations in the corresponding lattice gauge theories over a sufficient range of lattice spacings, and with enough precision that we can obtain plausible continuum extrapolations. We also put effort to extrapolate to the $N = \infty $ limit and to compare this to the physically interesting $SU(3)$ theory. To do so we have performed our calculations for $N = 2, 3, 4, 5, 6, 8, 10, 12$ gauge theories.



The structure of these proceedings is the following. First, in Section~\ref{sec:largeN} we provide a short chapter on the Large-$N$ limit to remind ourselves the basic properties of the physics on the planar limit.  Then in Section~\ref{sec:low_energy_effective_string_theory} we present the effective string theoretical descriptions suitable for approximating the spectrum of the confining string. Subsequently in Section~\ref{sec:lattice_calculation}, we provide a brief description of the lattice setup, by explaining how one can extract the masses of colour singlets on the lattice as well as the quantum numbers relevant for the extraction of the flux-tube and glueball spectra. Followingly, in Section~\ref{sec:results} we move to the presentation of the results starting from the spectra of confining strings, demonstrating the appearance of the worldsheet axion and proceeding to the spectrum of glueballs. Finally, in Section~\ref{sec:conclude}, we conclude.

\section{Large-$N$ limit}
\label{sec:largeN}
Yang-Mills gauge theory has a dimensionless running coupling $g^2$ and we, thus, might expect to be able to use the coupling as a general parameter for the theory. However, due to the fact that the scale invariance is anomalous, setting $g^2$ to some particular value $g^2_s$, we can only hope to use it as a useful expansion parameter for physics close to the scale $l_s$ for which the running coupling takes that value; in other words where $g^2(l=l_s) \simeq g^2_s$. 

\begin{figure}[h]
    \centering
    \vspace{0cm}
    \includegraphics[height=3cm]{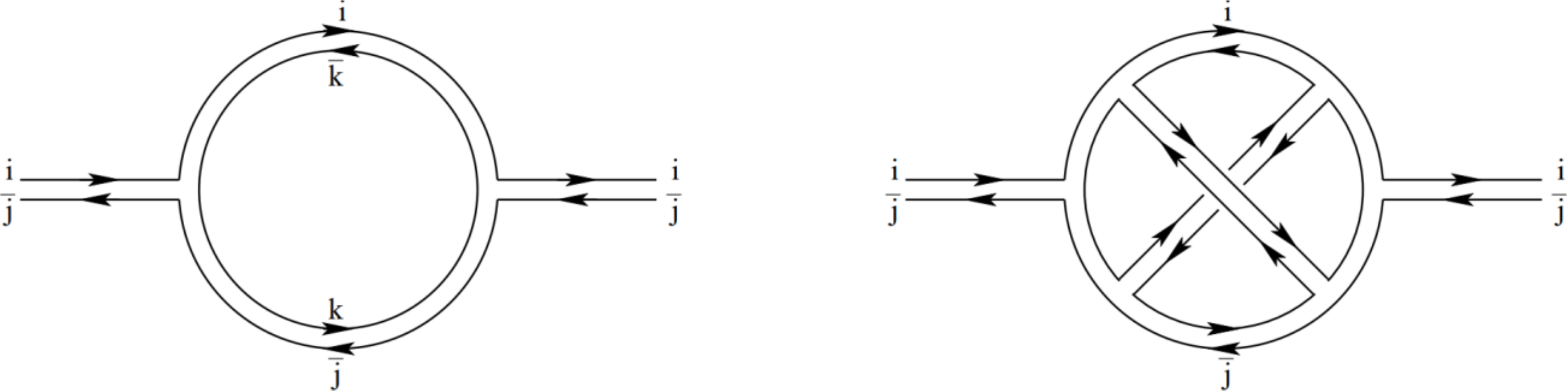}
     \vspace{0cm} 
    \caption{ \underline{Left Panel}: Example of a planar diagram, \underline{Right Panel}: Example of a non-planar diagram}
    \label{fig:largeNdiagrams}
\end{figure}

An alternative but more general expansion might be provided by $1/N$ as t' Hooft suggested, back in 1974. One can think of expanding $SU(N)$ gauge theories in powers of $1/N^2$ around $SU(\infty)$:
\begin{eqnarray}
 SU(N) = SU(\infty) + O(1/N^2).
\end{eqnarray}
According to the t' Hooft's double line representation for the gluon propagators and the associated vertices, ignoring for simplicity the difference between $U(N)$ and $SU(N)$, the expansion parameter can be expressed as $1/N^2$. As a result, a smooth large-$N$ limit can be achieved if one keeps the parameter $g^2 N$ fixed. This can be viewed by considering a gluon loop insertion in the gluon propagator using the double-line notation as this is pictorially represented on the left panel of Figure~\ref{fig:largeNdiagrams}.
The two vertices give a factor of $g^2$ and the sum over the colour index in the closed loop gives a factor of $N$. 
Hence, such an insertion will produce a factor of $g^2 N$ in the amplitude. To ensure smooth physics as increasing $N \to \infty$ we require that the number of such insertions in the diagrams dominating the physics of interest  are roughly fixed as we alter $N$. The above requires that we keep $\lambda = g^2 N$ fixed. Such  diagrams can be mapped on a plane and can, thus, be called planar. On the other hand diagrams on which glue propagators cross, cannot be mapped on a plane but can be resembled as planar diagrams with handles; an example of such a diagram is presented in the right panel of Figure~\ref{fig:largeNdiagrams}. This non-planar Feynman diagram has six vertices and just one circulating loop, which means that the expression will be proportional to $\sim N g^6 = \lambda^3 / N^2 \xrightarrow{N \to \infty} 0$. This is a naive way to demonstrate that the non-planar Feynman diagrams vanish in the large-$N$ limit. It is also straightforward to show that a Feynman diagram that contains virtual quark loops will get suppressed in the large-$N$ limit. Therefore, at the ’t Hooft limit only planar Feynman diagrams without quark loops survive.

So far we are making the assumption that there is a confining phase in the large-$N$ limit. This is based on numerical evidence. For instance, flux-tubes and glueballs exist and their masses extrapolate well in the Large-$N$ limit. We draw this conclusion by performing calculations for $SU(N)$ gauge theory and a sequence of finite values of $N$. Of course it would be nice to show that there is in fact a large-$N$ confining
phase, and that a smooth physics limit does in fact exist. 

At this point it should be made clear that there is no expectation that all the physics of $SU(3)$ is close to that of $SU(\infty)$. We can only make sure that an observable in $SU(3)$ is close to that of $SU(\infty)$ once we perform the calculation. It could be possible that other large-$N$ limits are more appropriate for the physics under investigation. For instance in QCD where we have 2 or 3 light flavours, $N_f/N \sim 1$. Hence, it could appear possible that the limit $N \to \infty$ by keeping $N_f/N$ fixed might be more appropriate for some physical quantities~\cite{Veneziano:1976wm}. A nice review where several such limits are being discuss is provided in Ref.~\cite{Manohar:1998xv}

\section{Low energy Effective String Theories}
\label{sec:low_energy_effective_string_theory}
Let us imagine a flux-tube as a confining string with length $l=aL_{x}$ winding around the spatial torus where $a$ the lattice spacing. Imposing fixed spatial position for the string  spontaneously  breaks  translation  symmetry. Therefore, we expect $D-2$ Nambu-Goldstone massless bosons to appear at low energies. Such bosons reflect the transverse fluctuations of the flux-tube around its classical configuration. We would thus, expect a low energy Effective String Theory (EST) describing the flux-tube spectrum for large enough strings. Of course a flux-tube is not an infinitesimally thin string, it is an $SU(N)$ object and presumably has an intrinsic width $l_w\propto 1/\sqrt{\sigma}$. We would therefore expect that the spectrum of the flux-tube consists not only of string like states but also of massive excitations. Below, we describe the current theoretical predictions for the excitation spectrum of the Nambu-Goldston bosons as well as an approach to explain the existence of massive resonances on the world-sheet of the confining string.
\subsection{The Goddard–Goldstone–Rebbi–Thorn string}
\label{sec:GGRT}
In this subsection we describe the spectrum of the Goddard-Goldstone-Rebbi-Thorn (GGRT)~\cite{Goddard} string or in other words the Nambu-Goto (NG)~\cite{NGpapers} closed string. NG string describes non-critical relativistic bosonic strings. One can extract the GGRT spectrum by performing light-cone quantization of the closed-string using the NG action or equivalently the Polyakov action. This model is Lorentz invariant only in $D=26$ dimensions. Nevertheless, for reasons that we now understand better~\cite{Dubovsky:2013gi} NG can also describe to a good extend the spectrum of strings in $D=3 \ {\rm and} \ 4$ dimensions. The expression of the GGRT spectrum is given by:
\begin{equation}
{E_{N_L,N_R}(q,l)}
= \sigma l \sqrt{
1 
+
\frac{8\pi}{(l \sqrt{\sigma})^2} \left(\frac{N_L+N_R}{2}-\frac{D-2}{24}\right)
+
\left(\frac{2\pi q}{(l\sqrt{\sigma})^2}\right)^2}\,,
\label{eqn_EnNG}
\end{equation}
where $2\pi N_{L(R)}/l$ the total energy and momentum of the left(right) moving phonons with $N_L = \sum_k \sum_{n_L(k)} k(n^+_L(k)+ n^-_L(k))$ and
 $N_R = \sum_k \sum_{n_R(k)} k(n^+_R(k)+ n^-_R(k))$. $n^{\pm}_{L(R)}(k)$ is the number 
of left(right) moving phonons of momentum $p_k = 2\pi k/l$, $k=0,1,2,\dots$ and angular momentum  $\pm 1$. If $p_{||}=2\pi q/l$ is the total longitudinal momentum of the string then, since the phonons provide that momentum, we must have $N_L - N_R = q$ (level matching constrain). The angular momentum (spin) around the string is expressed as $J = \sum_{k,n_L(k),n_R(k)} n^+_L(k) + n^+_R(k) - n^-_L(k) - n^-_R(k)$.


\subsection{Lorentz invariant string approaches}
\label{sec:Lorentz}
 Systematic ways to study Lorentz invariant EST which can describe the confining string were pioneered by  L\"uscher, Symanzik, and Weisz in \cite{Luscher} (static gauge) as well as by Polchinski and Strominger in~\cite{Pol} (conformal gauge). Such approaches produce predictions for the energy of states as an expansion in $1/l\sqrt{\sigma}$. Terms in this expansion of $O(1/l^p)$ are generated by $(p+1)$~-~derivative terms in the EST action whose coefficients are a priori arbitrary Low Energy Coefficients (LECs). Interestingly, these LECs were shown to obey strong constraints that reflect a non-linear realization of Lorentz symmetry \cite{LW,Meyer,AK}, and so to give parameter free predictions for certain terms in the $1/l$ expansion.
 
 The EST approaches can be characterised by the way one performs the gauge fixing of the embedding coordinates on the world-sheet. This can be either the static gauge~\cite{Luscher,LW,AK} or the conformal gauge~\cite{Pol,Drummond:2004yp,HariDass:2009ub} with both routes leading to the same results. The starting point of building the EST is the leading area term which gives rise to the linearly rising potential for large strings i.e. $E \simeq \sigma l$. Subsequently comes the Gaussian action which is responsible for the $\propto 1/l$ L\"uscher term with universal coefficient depending only on the dimension $D$. At next step one adds the 4-derivative terms which yield a correction on the energy spectrum proportional to $1/l^3$ with a universal coefficient that also depends on the dimension $D$. One can include the $6-$derivative terms and show that for $D=3$ they yield the fourth universal term proportional to $1/l^5$ in the energy spectrum, while for general states in $D=4$, the coefficient of the $O(1/l^5)$ term is not universal. Nonetheless, the energy just for the ground state in the $D=4$ case is universal. Summarizing the above information, the spectrum is given by 
 \begin{eqnarray}
  E_n(l) =  \sigma l + \frac{4 \pi}{l}\bigg(n-\frac{D-2}{24}\bigg)  -  \frac{8 \pi^2}{\sigma l^3}\bigg(n-\frac{D-2}{24}\bigg)^2  +   \frac{32 \pi^3}{\sigma^2 l^5}\bigg(n-\frac{D-2}{24}\bigg)^3   + {O}(l^{-7}).
\label{eq:AharonyKarzbrun}
\end{eqnarray}
Since we think of the GGRT model as an EST, which may be justified only for long strings \cite{Olesen}, one can expand the associated energy for $l\sqrt{\sigma}\gg 1$. The result of the expansion is the same as the expression in Equation~\ref{eq:AharonyKarzbrun} where for simplicity we have set $q=0$, and  $n = (N_L + N_R)/2$. 

\subsection{The topological term action}
\label{sec:axion}
In 2013, Dubovsky {\it et al.} worked out an approach for extracting the spectrum of the confining string for short as well as for long lengths. The idea was based on the fact that the GGRT string provides the best approximation for the flux-tube spectrum and that Equation~\ref{eqn_EnNG} can be re-expressed as $E_{N_L,N_R} = \sqrt{\sigma} {\cal E} (p_k/\sqrt{\sigma},1/l \sqrt{\sigma})$ where $p_k$ are the momenta of individual phonons in units of $2 \pi / l$ comprising the state quantised. The naive expansion in terms of $1/ l \sqrt{\sigma}$ is the combination of two different expansions; the first is an expansion in the softness of individual quanta compared to the string scale, i.e. in $p_k / \sqrt{\sigma}$ and the second expansion is a large volume expansion, i.e. an expansion in $1 / l \sqrt{\sigma}$. To disentangle the two expansions the following procedure is being adopted. First, one calculates the infinite volume $S$-matrix of the phonon collisions. This is done perturbativelly given that the center of mass energy of the colliding phonons is small in string units; this is called the momentum expansion. Followingly, the authors extracted the finite volume energies from this $S$-matrix by using approximate integrability and the Thermodynamic Bethe Ansatz (TBA). This allows to extract the winding effects on the energy from virtual quanta traveling around the circle as well as the winding corrections due to phonon interactions.

The authors argued that when a state has only left-moving phonons the GGRT winding corrections in the string spectrum are small and, therefore, one expects the spectrum to be close to that of the free theory. On the contrary, for states containing both left- and right-moving phonons, energy corrections are larger. The above is in a good agreement with most of the states in $D=4$ but fails to provide an explanation for the anomalous behaviour of the pseudoscalar level $0^{--}$ firstly demonstrated in \cite{Athenodorou:2010cs, Athenodorou:2009Sept}, suggesting that an additional action term is required in order to describe such excitations. The most straightforward way to do this is the introduction of a massive pseudoscalar particle $\varphi$ on the world-sheet. The leading interaction compatible with non-linearly realized Lorentz
invariance for such a state is a coupling to the topological invariant known as the self-intersection number of the string $S_{\rm int} = \frac{\alpha}{8 \pi} \int d^2 \sigma \varphi K^{i}_{\alpha \gamma} K^{j \gamma}_{\beta} \epsilon^{\alpha \beta} \epsilon_{ij} \,$
with $K^{i}_{\alpha \gamma}$ being the extrinsic curvature of the world-sheet, $\alpha$ the associated coupling and $\sigma^{i}$, $i=1,2$ the world-sheet coordinates. Adapting the above interaction term to our old results for $SU(3)$, $\beta=6.0625$ yields a mass of $m_{\varphi}/\sqrt{\sigma} \simeq 1.85^{+0.02}_{-0.03}$ and a coupling of $\alpha = 9.6  \pm 0.1$.

\vspace{-0.25cm}
\section{Lattice calculation}
\label{sec:lattice_calculation}
\subsection{The lattice gauge theory}
\label{sec:lattce_gauge_theory}
We define the $SU(N)$ gauge theory on a $D=4$ Euclidean space-time lattice which has been compactified along all directions with volume $L_{x} \times L_{y} \times L_{z} \times L_{T}$. The length of the flux-tube is equal to $L_{x}$, while $L_{y}$, $L_{z}$ and $L_{T}$ were chosen to be large enough to avoid finite volume effects. For the calculation of the confining string spectra we choose the transverse lattice extents $L_{y} = L_{z}  = L_{\perp}$ uniformly so that we ensure rotational symmetry around the string axis while for glueballs we choose all spatial directions to be equal i.e. $L_{x} = L_{y} = L_{z}$ for similar reasons. We perform Monte-Carlo simulations using the standard Wilson plaquette action
\begin{eqnarray}
 S=\sum_{\square} \beta \left[ 1-\frac{1}{N}{\rm Re}{\rm Tr}(U_{\square}) \right]\,,
\end{eqnarray}
where the sum runs over all the plaquettes ($\square$), the basic square Wilson loop one can construct with side one lattice spacing $a$ as well as with inverse coupling $\beta=\frac{2N}{g^2(a)}$. In order to keep the value of the lattice spacing $a$ approximately fixed for different values of $N$ we keep
the 't Hooft coupling $\lambda(a)=N g^2(a)$ approximately fixed, so that $\beta \propto N^2$. From a technical point of view, the simulation algorithm materialized for such investigations, combines standard heat-bath and over-relaxation steps in the ratio 1:4; these are implemented by updating $SU(2)$ subgroups using the Cabibbo-Marinari algorithm~\cite{Cabibbo:1982zn}. 

\subsection{Mass Extraction and Quantum Numbers}
\label{sec:mass}

Flux-tubes and glueballs are colour singlet states. Masses of colour singlet states can be calculated using the standard decomposition of a Euclidean correlator of some operator $\phi(t)$, with high enough overlap onto the physical states in terms of the energy eigenstates of the Hamiltonian of the system $H$:
\begin{eqnarray}
\langle \phi^\dagger(t=an_t)\phi(0) \rangle
& = &
\langle \phi^\dagger e^{-Han_t} \phi \rangle
=
\sum_i |c_i|^2 e^{-aE_in_t} \nonumber \\
& \stackrel{t\to \infty}{=} & 
|c_0|^2 e^{-aE_0n_t}\,,
\label{extract_mass}
\end{eqnarray}
where the energy levels are ordered, $E_{i+1}\geq E_i$, with  $E_0$ that of the ground state. The only states that contribute in the above summation are those that have non zero overlaps i.e. $c_i = \langle {\rm vac} | \phi^\dagger | i \rangle \neq 0$. We, therefore, need to match the quantum numbers of the operator $\phi$ to those of the state we are interested in. In this work we are interested in glueballs and closed flux-tubes, thus, we need to encode the right quantum properties within the operator $\phi$ which will enable us to project onto the aforementioned states.

The extraction of the ground state relies on how good the overlap is onto this state and how fast in $t$ we obtain the exponential decay according to Eq.~(\ref{extract_mass}). The overlap can be maximized by building operator(s) which "capture" the right properties of the state, in other words by projecting onto the right quantum numbers as well as onto the physical length scales of the relevant state. In order to achieve a decay behaviour setting in at low values of $t$ one has to minimize contributions from excited states. To this purpose we employ the variational calculation or GEVP (Generalized Eigenvalue Problem)~\cite{Luscher:1984is,Luscher:1990ck} applied to a basis of operators built by several lattice path in different blocking levels \cite{variational,MT-block,BLMTUW-2004}. This reduces the contamination of excitation states onto the ground state and maximizes the overlap of the operators onto the physical length scales.

\subsubsection{Quantum numbers of the confining string}
\label{sec:quantum_numbers_confining_string}

The energy states of the closed flux-tube in $D=3+1$ are characterised by the irreducible representations of the two-dimensional lattice rotation symmetry around the principal axis denoted by $C_4$~\cite{Kuti1}. The above group is a subgroup of $O(2)$ corresponding to rotations by integer multiples of $\pi/2$ around the flux-tube propagation axis. This splits the Hilbert space in four orthogonal sectors, namely: $J_{{\rm mod}\, 4}=0$, $J_{{\rm mod}\, 4}=\pm 1$,  $J_{{\rm mod}\, 4}=2$. Furthermore, parity $P_\perp$ which is associated with reflections around the axis $\hat{\perp}_1$ can be used to characterise the states. Applying $P_\perp$ transformations, flips the sign of $J$. Therefore, one can choose a basis in which states are characterised by their value of $J$ ($J= \pm$), or by their value of $|J|$ and  $P_\perp$. We adopt the latter. In the continuum, states with $J \neq 0$ are parity degenerate, however, on the lattice this holds only for the odd values of $J$. In practice, we describe our states with the following $5$ irreducible representations $A_{1}$, $A_{2}$, $E$, $B_{1}$ and $B_{2}$ of $C_4$ group whose $J$ and $P_\perp$ assignments are: $\left\{ A_1: \, |J_{{\rm mod} \ 4}|=0, \ P_\perp=+\right\}, \left\{A_2:  \, |J_{{\rm mod} \ 4}|=0, \  P_\perp=-\right\}$, $\left\{ E: |J_{{\rm mod} \ 4}|=1, \  P_\perp=\pm\right\}$,  $\left\{ B_1: \, |J_{{\rm mod} \ 4}|=2, \ P_\perp=+\right\}$ and $\left\{B_2:  \, |J_{{\rm mod} \ 4}|=2, \ P_\perp=-\right\}$.

Furthermore, there is the longitudinal momentum $p_{||}$ carried by the confining string along its axis (which is quantized in the form $p_{||}=2\pi q/L_x; q\in Z$) and the parity $P_{||}$ with respect to reflections across the string midpoint. Since $P_{||}$ and $p_{||}$ do not commute, we can use both to  simultaneously characterise a state only when $q=0$. The energy does not depend on the sign of momentum $q$ and we, thus, focused on those with $q\ge 0$.
\begin{figure}[h]
    \centering
    \vspace{0cm}
    \rotatebox{0}{\includegraphics[height=7cm]{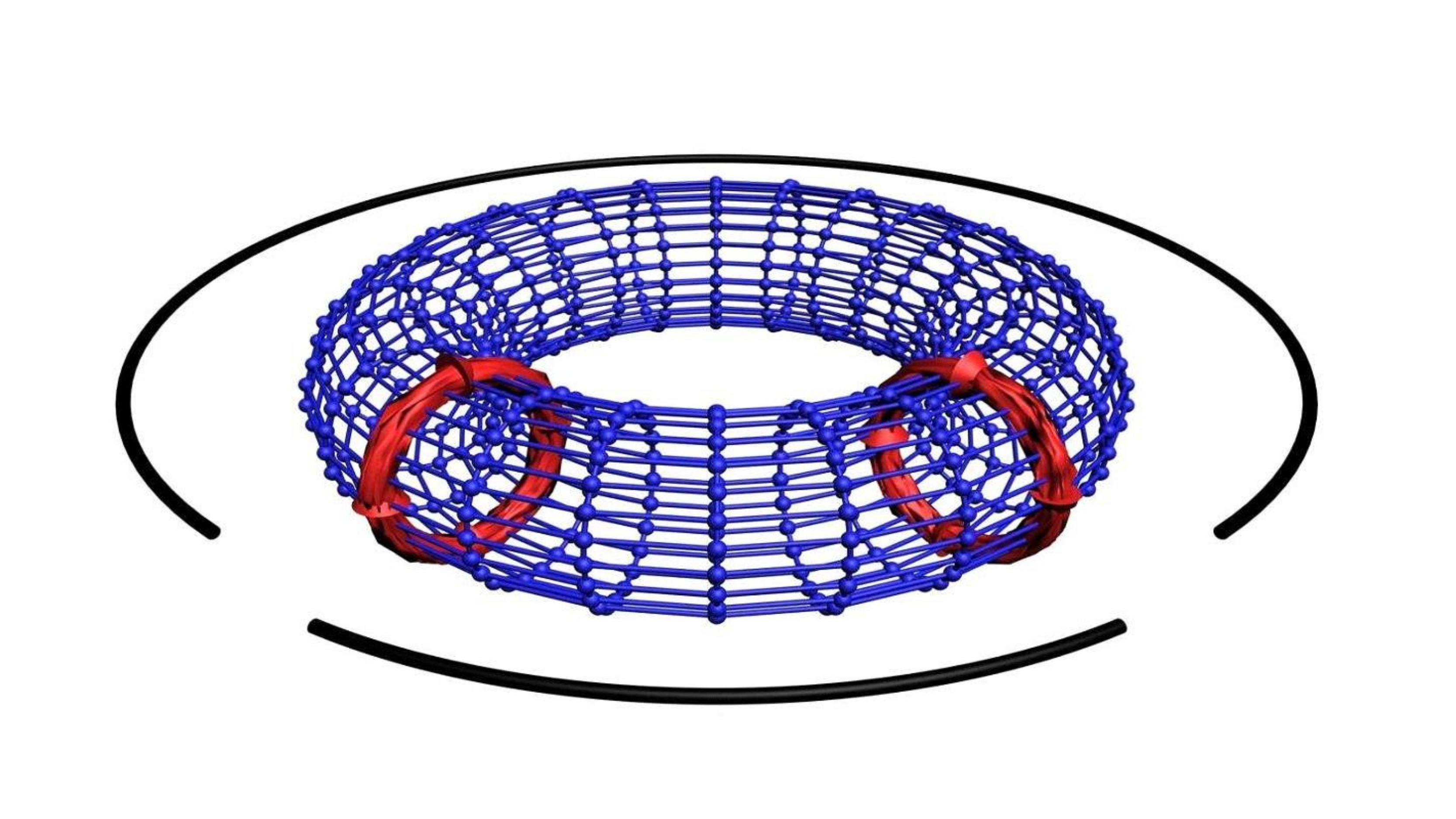}\put(-197,180){\LARGE $T-t$}\put(-182,0){\LARGE $t$}\put(-300,50){\LARGE $\phi^{\dagger}_i$}\put(-77,50){\LARGE $\phi_j$}}
    \caption{A pictorialisation of the torelon correlation function in $D=2$ Euclidean dimensions.}
    \label{fig:torelon_correlator}
\end{figure}
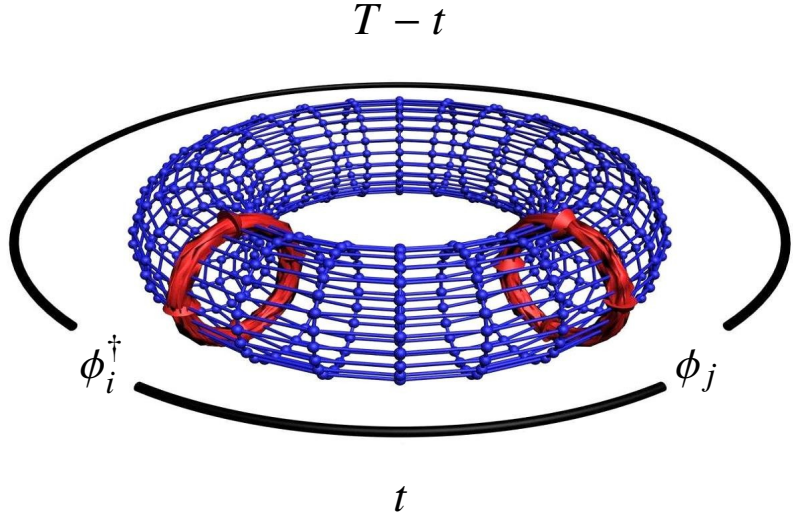

\begin{figure}
    \centering
    \includegraphics[height=8.75cm]{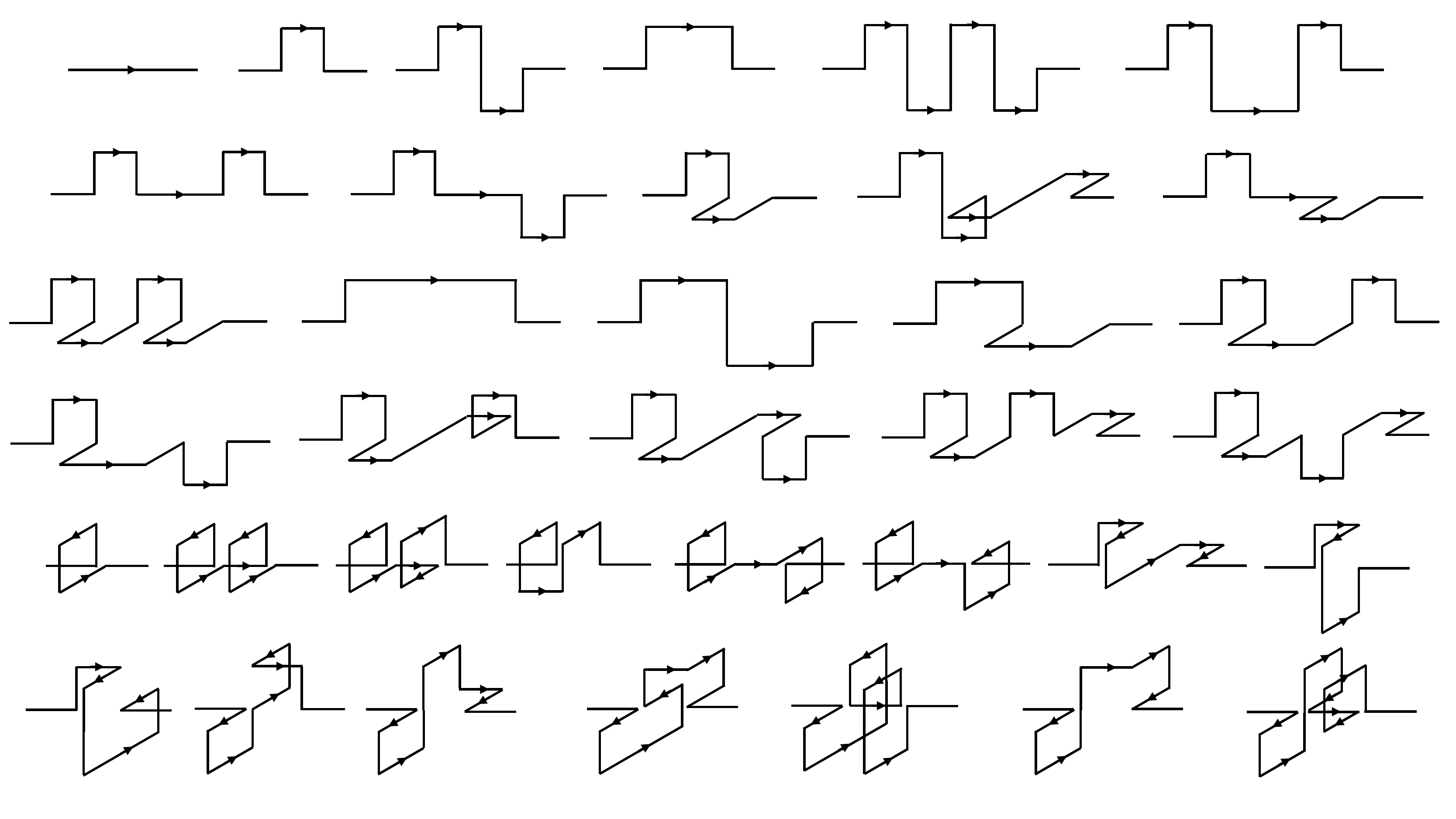}
    \caption{All the different paths used for the construction of the torelon operators.}
    \label{fig:paths}
\end{figure}

Flux-tube energies are extracted by making use of correlation matrices $C_{ij} = \langle \phi_i^{\dagger} (t) \phi_j (0) \rangle$ with $i,j=1...N_{\rm op}$ in combination with GEVP where $N_{\rm op}$ the number of operators. A pictorialisation of a correlation function between closed flux-tube operators is provided in Figure~\ref{fig:torelon_correlator}. We construct operators $\phi_i$ which encode shapes that lead to particular values of $J,P_\perp,P_{||},$ and $q$. We do so by choosing 
linear combinations of Polyakov loops the paths of which consist of  various  transverse deformations and various smearing and blocking levels~\cite{variational}. All the transverse paths used for the construction of the operators are shown in Figure~\ref{fig:paths} and all together, including smearing and blocking levels, form a basis of around $N_{\rm op} =1000$ operators with approximately $50 - 200$ for each different irreducible representation. To build an operator which encodes a certain value of angular momentum $J_{{\rm mod} \ 4}$ we begin the construction with a sub-operator $\phi_{\alpha}$ which has a deformation extending in angle $\alpha$ within the plane of transverse directions. Then we repeat the same procedure by rotating the sub-operator by integer values of $\pi/2$. Finally, we can construct the operator $\phi(J)$ belonging to a specific representation of $C_{4}$ by using the formula $\phi(J)= \sum_{n=1,2,3,4} e^{iJ n \frac{\pi}{2}} \phi_{n \frac{\pi}{2}}\,.$ Thus $\phi(0)$ belongs to $A_1$ and $A_2$, $\phi(1)$ to $E$ and, finally, $\phi(2)$ to $B_1$ as well as $B_2$. Lastly, it is required to encode certain values of $P_{\perp}$ and $P_{||}$ by summing and subtracting reflections of the initial sub-operator $\phi(J)$ over the transverse and parallel parity planes. Such an example is pictorialized in Equation~\ref{eq:example} for an operator with $J_{\rm mod \ 4}=0$.
\begin{eqnarray}
 \phi= {\rm Tr}\left[ \parbox{12.5cm}{\rotatebox{0}{\includegraphics[width=12.5cm]{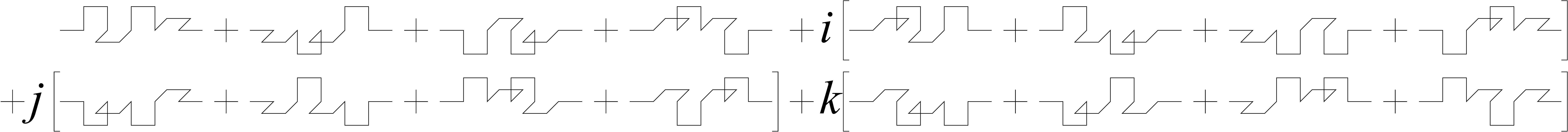}}} \ \right]\,.
\label{eq:example}
\end{eqnarray}
For the combination $i=j=k=+1$, $\phi$ projects onto $\{ A_1, P_{||}=+ \}$, for $i=+1,j=k=-1$ onto $\{ A_2, P_{||}=+ \}$, for $i=-1,j=+1,k=-1$ onto $\{ A_1, P_{||}=- \}$ and finally, for $i=j=-1,k=+1$, onto $\{ A_2, P_{||}=- \}$.

\subsubsection{The Quantum Numbers of glueballs}
\label{sec:quantum_numbers_glueballs}

\begin{figure}[h]
    \centering
    \includegraphics[height=9cm]{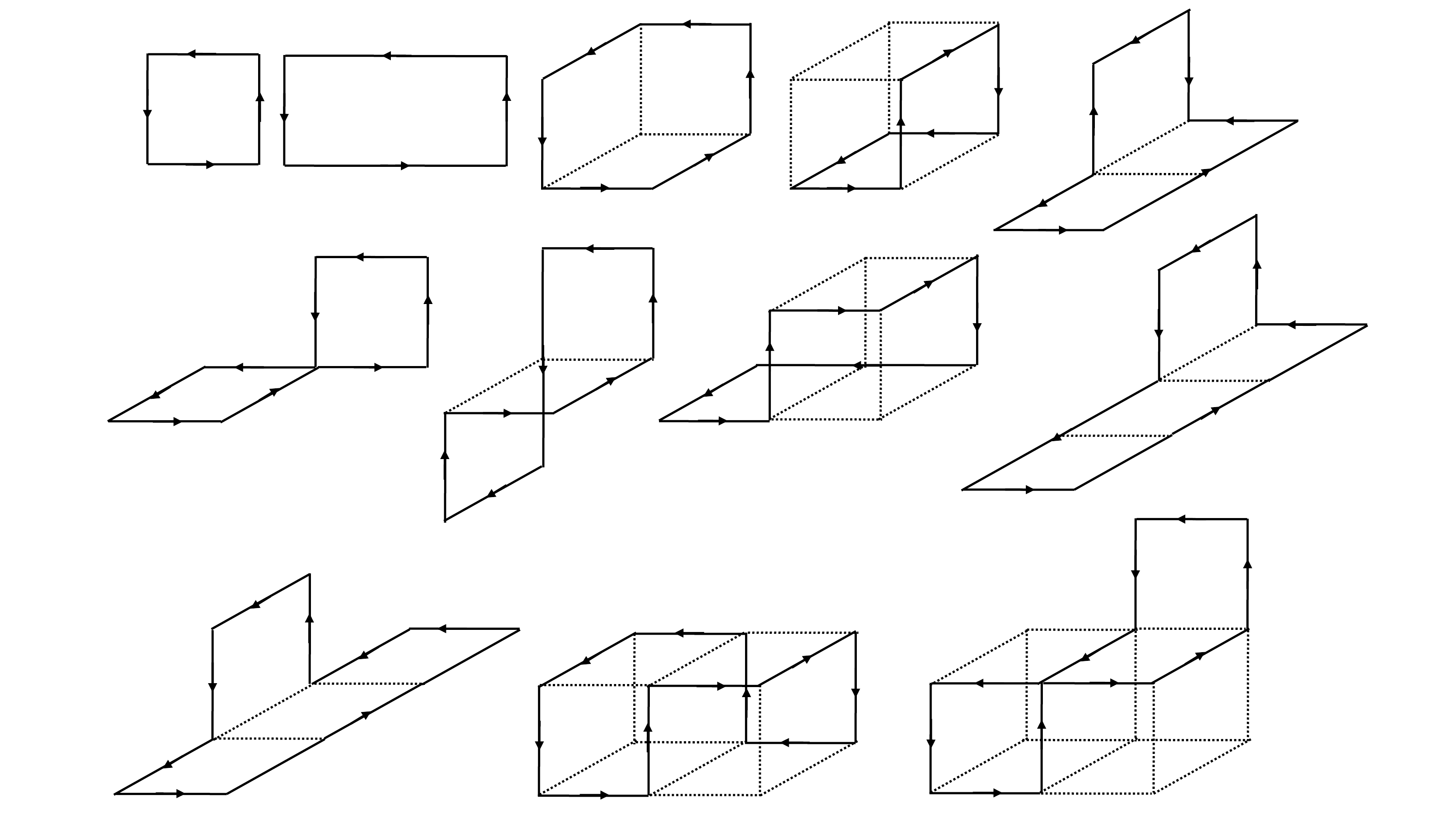}
    \caption{The 12 different closed loops used for the construction of the glueball operators.}
    \label{fig:glueball_operators}
\end{figure}

The glueballs like the flux-tubes are color singlets and, thus, an operator projecting onto a glueball state is obtained by taking the ordered product of $SU(N)$ link matrices however now around a contractible loop and then taking the trace. To retain the exact positivity of the correlators we use loops that contain only spatial links. The real part of the trace projects on $C = +$ and the imaginary part on $C = -$. We sum all
spatial translations of the loop so as to obtain an operator with momentum $p=0$. We take all rotations of the loop and construct the linear combinations that transform according to the irreducible representations, $R$, of the rotational symmetry group of our cubic spatial lattice.
We always choose to use a cubic spatial lattice volume ($L_x=L_y=L_z$) that respects these symmetries. For each
loop we also construct its parity inverse so that taking linear combinations we can construct
operators of both parities, $P = \pm$. The correlators of such operators will project onto glueballs
with $p = 0$ and the $R^{P C}$ quantum numbers of the operators concerned. All the 12 paths used for the construction of the glueball operators are provided in Figure~\ref{fig:glueball_operators}.

The irreducible representations $R$ of our subgroup of the full rotation group are usually
labelled as $A_1,A_2,E,T_1,T_2$. Mind that these representations are different than those for the group $C_4$ of the confining-string . The $A_1$ is a singlet and rotationally symmetric, so it
will contain the $J=0$ state in the continuum limit. The $A_2$ is also a singlet, while the
$E$ is a doublet and $T_1$ and $T_2$ are both triplets. Since, for example, the three states transforming as the triplet of $T_2$ are degenerate on
the lattice, we average their values and treat them as one state in our estimates of glueball
masses and we do the same with the $T_1$ triplets and the $E$ doublets.

Once more, the glueball energies are extracted by making use of correlation matrices $C_{ij} = \langle \phi_i^{\dagger} (t) \phi_j (0) \rangle$ with $i,j=1...N_{\rm op}$ in combination with GEVP where $N_{\rm op}$ the number of operators. The scalar channel $A_1^{++}$ has a non-zero projection onto the vacuum.  In this case it can be convenient to use the vacuum-subtracted operator $\phi_i - \langle \phi_i \rangle$, which will remove the contribution of the vacuum in Equation~\ref{extract_mass}, so that the lightest
non-trivial state appearing in the aforementioned sum, is the leading term in the expansion of states.

The above representations of the rotational symmetry reflect our cubic lattice formulation. As we approach the continuum limit these
states will approach the continuum glueball states which belong to representations of the continuum rotational symmetry. In other words they fall into degenerate multiplets of $2J + 1$ states. In determining the continuum limit of the low-lying glueball
spectrum, it is clearly more useful to be able to assign the states to a given spin $J$, rather
than to the representations of the cubic subgroup which have a much less fine ‘resolution’
since all of $J =1,2,3 \dots, \infty$,  are mapped to just 5 cubic representations. The way $2J + 1$
states for a given $J$ are distributed amongst the representations of the cubic symmetry subgroup is given, for the relevant low values of $J$, in table~\ref{tab:table_J_R}. For instance, the seven states corresponding to a $J = 3$ glueball will be distributed over a singlet $A_2$, a degenerate triplet $T_1$ and a degenerate triplet $T_2$, so seven states in total. These $A_2$, $T_1$ and $T_2$ states will be split by $O(a^2)$ lattice spacing corrections. So once the lattice spacing $a$ is small enough these states will be nearly degenerate and one can use this near-degeneracy to identify the continuum spin. 

\begin{table}[h]
\centering
\begin{tabular}{ccc} \hline\hline
\multicolumn{3}{c}{continuum $J \sim$ cubic $R$} \\ \hline\hline
$J$    &        &  cubic $R$  \\  \hline \hline
 0   & $\sim$ & $A_1$   \\
 1   & $\sim$ & $T_1$      \\
 2   & $\sim$ & $E+T_2$     \\
 3   & $\sim$ & $A_2+T_1+T_2$     \\
 4   & $\sim$ & $A_1+E+T_1+T_2$     \\
 5   & $\sim$ & $E+2T_1+T_2$      \\
 6   & $\sim$ & $A_1+A_2+E+T_1+2T_2$     \\
 7   & $\sim$ & $A_2+E+2T_1+2T_2$     \\
 8   & $\sim$ & $A_1+2E+2T_1+2T_2$     \\ \hline \hline
\end{tabular}
\caption{Projection of continuum spin $J$ states onto the cubic representations $R$.}
\label{tab:table_J_R}
\end{table}

\section{Results}
\label{sec:results}

\subsection{The spectrum of the confining string and the world-sheet axion}
\label{sec:results_cofining_string}

At this section of the manuscript we present results for the spectrum of the confining string extracted from calculations on five different gauge groups. Namely, we investigated
$N=3$ at $\beta=6.0625$ ($a\simeq 0.09 {\rm fm}$) and $\beta=6.338$ ($a\simeq 0.06 {\rm fm}$), $N=5$ at $\beta=17.630$ ($a\simeq 0.09 {\rm fm}$) and $\beta=18.375$ ($a\simeq 0.06 {\rm fm}$) as well as $N=6$ at  $\beta=25.550$ ($a\simeq 0.09 {\rm fm}$). Critical slowing down~\cite{Athenodorou:2021qvs,Athenodorou:2020ani}, as one moves towards the continuum ($a \to 0$) and the large-$N$ limit ($N \to \infty$), prohibits the investigation of gauge groups with $N \geq 6$ and $a < 0.09 {\rm fm}$. Nevertheless, the above configuration of measurements is enough to determine whether significant lattice artifacts as well as $1/N^2$ corrections are affecting our statistically more accurate $N=3$ calculations. As a matter of fact our investigation demonstrates that such effects are of minor importance and do not play a significant role in the interpretation of the spectrum. The energy spectrum we extracted is compared to the predictions of the GGRT string. Namely, we fit the absolute ground state ($|J_{\rm mod \ 4}|^{P_{\perp} P_{||}} =0^{++}$) for all calculations using Equation~\ref{eqn_EnNG} as a function of the length for $l\sqrt{\sigma} > 2.5$ and extract the string tension $a \sqrt{\sigma}$. Once the string tension has been extracted, Equation \ref{eqn_EnNG} can be used as a parameter free prediction for higher string excitations with $N_L + N_R > 0$. 

\subsubsection{The energy spectrum for $q=0$ and the world-sheet axion}
\label{sec:resultsq0}
\begin{figure}[h]
    \centering  
    \vspace{-1.0cm}
    \includegraphics[height=10cm]{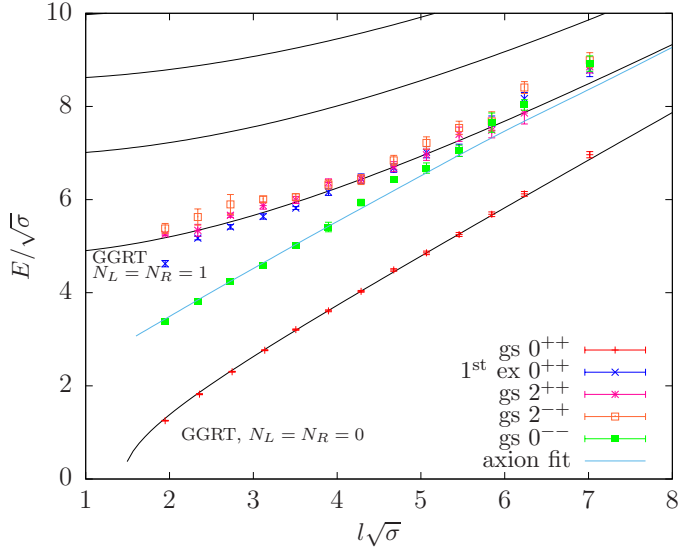} 
    \vspace{-1cm}
    \caption{The spectrum of the absolute ground state and first excited state for $|J_{\rm mod \ 4}|^{P_{\perp}, P_{||}}=0^{++}$ as well as the ground states for $2^{++}$, $2^{-+}$ and the "anomalous"  $0^{--}$ for $q=0$ and $SU(3)$ at $\beta=6.0625$; the black lines correspond to the GGRT predictions and the light blue line to the prediction of the EFT with the axionic part of the action included within.}
    \label{fig:first_1}
\end{figure}

\begin{figure}[h]
    \centering
    \vspace{-1.0cm}
    \includegraphics[height=10cm]{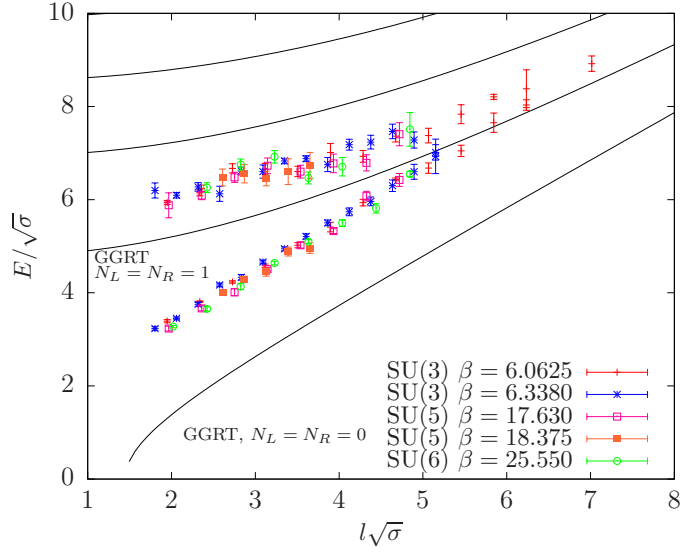} 
    \vspace{-1cm}
    \caption{The energies of the ground state and first excited state for $|J_{\rm mod \ 4}|^{P_{\perp}, P_{||}}=0^{--}$, $q=0$ for all gauge groups considered in this work.}
    \label{fig:first_2}
\end{figure}

\begin{figure}[h]
    \centering
        \vspace{-1.0cm} 
    \includegraphics[height=10.0cm]{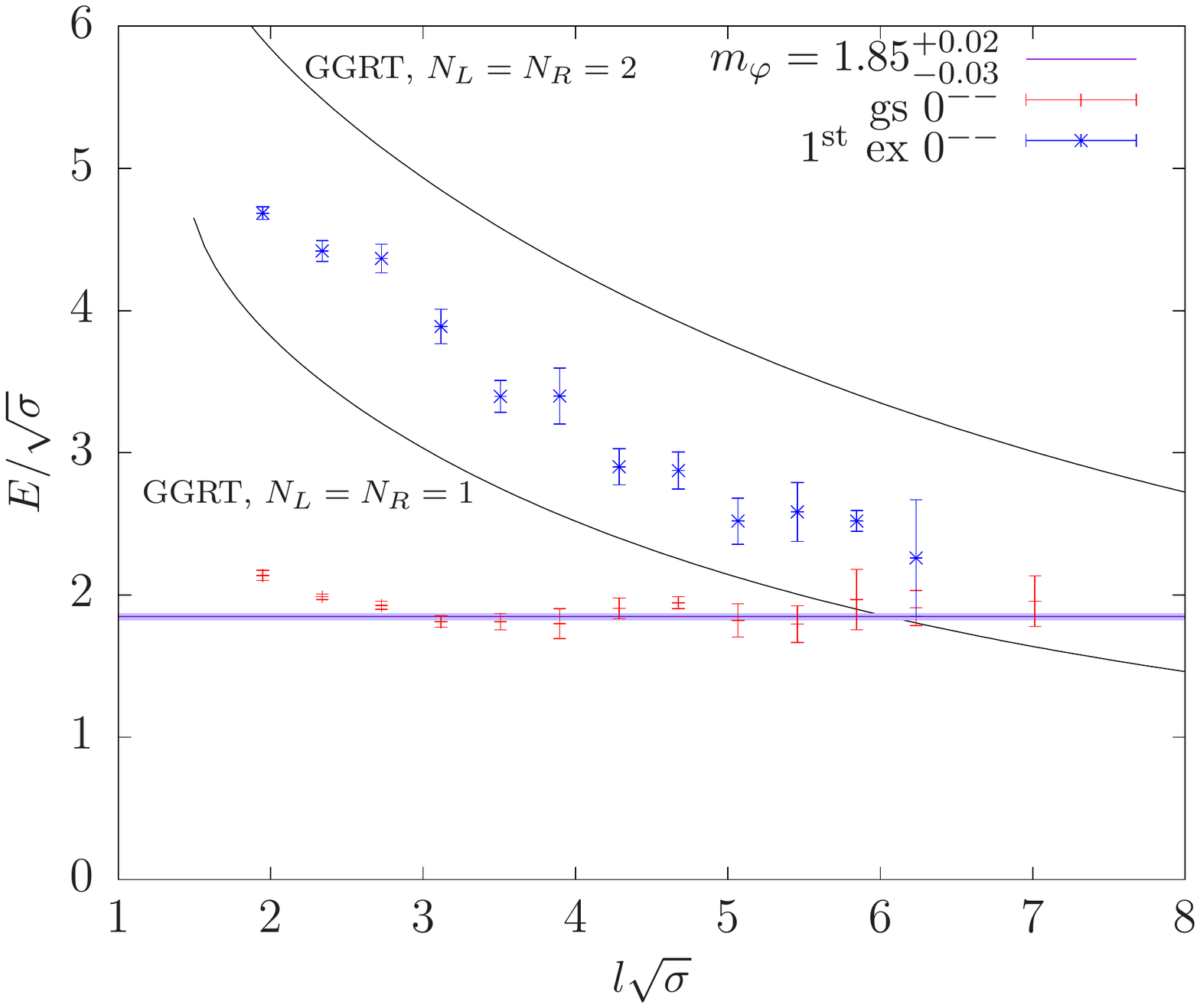}
    \vspace{-1cm}
    \caption{The energy levels of the ground and first excited states for a closed flux-tube with quantum numbers $0^{--}$, $q=0$ and the zeroth energies subtracted for $SU(3)$, $\beta=6.0625$. The horizontal purple band corresponds to the mass of the axion as this has been extracted in \cite{Dubovsky:2013gi}.}
    \label{fig:second}
\end{figure}

We begin by presenting our results for the $q=0$ longitudinal momentum sector in Figures~\ref{fig:first_1}, \ref{fig:first_2}, \ref{fig:second}, \ref{fig:third_1} and \ref{fig:third_2}. In Figure~\ref{fig:first_1}, the lowest energy level corresponds to the absolute ground state $|J_{\rm mod \ 4}|^{P_{\perp} P_{||}} =0^{++}$ which is used to set the scale of the NG string, hence, the nearly perfect agreement with the GGRT string. Furthermore in Figure~\ref{fig:first_1}, we plot the first excited state of $0^{++}$ as well as the ground states of $2^{++}$, $2^{-+}$ and $0^{--}$ for $SU(3)$ at $\beta=6.0625$. We compare the above data with the GGRT prediction for $N_R=N_L=1$. This string state is expected to be four-fold degenerate with levels with continuum quantum numbers $0^{++}$, $0^{--}$, $2^{++}$ and $2^{-+}$ . While $0^{++}$, $2^{++}$ and $2^{-+}$ flux-tube excitations appear to exhibit small deviations for short values of $l\sqrt{\sigma}$ and for larger strings become consistent with GGRT, $0^{--}$ ground state appears to demonstrate significant deviations from the GGRT string. In Figure~\ref{fig:first_2} we present the ground and in addition the first excited state with quantum numbers $0^{--}$ for all gauge groups considered in this work. It appears that both states are only mildly affected by lattice artifacts and $1/N^2$ corrections. The $0^{--}$ ground state appears to have characteristics of a resonance i.e. a constant mass term coupled to the absolute ground state. This is more obvious by subtracting the absolute ground state $0^{++}$ where this excitation exhibits a plateau; this  is presented in Figure~\ref{fig:second} for $SU(3)$ at $\beta=6.0625$. As has already being explained in Section~\ref{sec:axion} this state can be well interpreted as an axion on the world-sheet of the flux-tube with an associated mass of $m_{\varphi}/\sqrt{\sigma} = 1.85^{+0.02}_{-0.03}$ for $SU(3)$ at $\beta=6.0625$; This value is in good agreement with the plateau in Figure~\ref{fig:second}. If the $0^{--}$ flux-tube ground state corresponds to the axion, the next excitation level would correspond to the string-like state with $N_L=N_R=1$ rather than  $N_L=N_R=2$. As one can see in the right panel of Figure~\ref{fig:first_1},  and Figure~\ref{fig:second} the $0^{--}$ first excitation state does not approach the GGRT $N_L=N_R=2$ state but instead it slowly approaches the $N_L=N_R=1$ string state. This strengthens the scenario of $0^{--}$ ground state being 
the world-sheet axion.

\subsubsection{A state of two axions}
\label{sec:two_axions}
\begin{figure}[h]
    \centering
    \vspace{-1.0cm}
    \includegraphics[height=10cm]{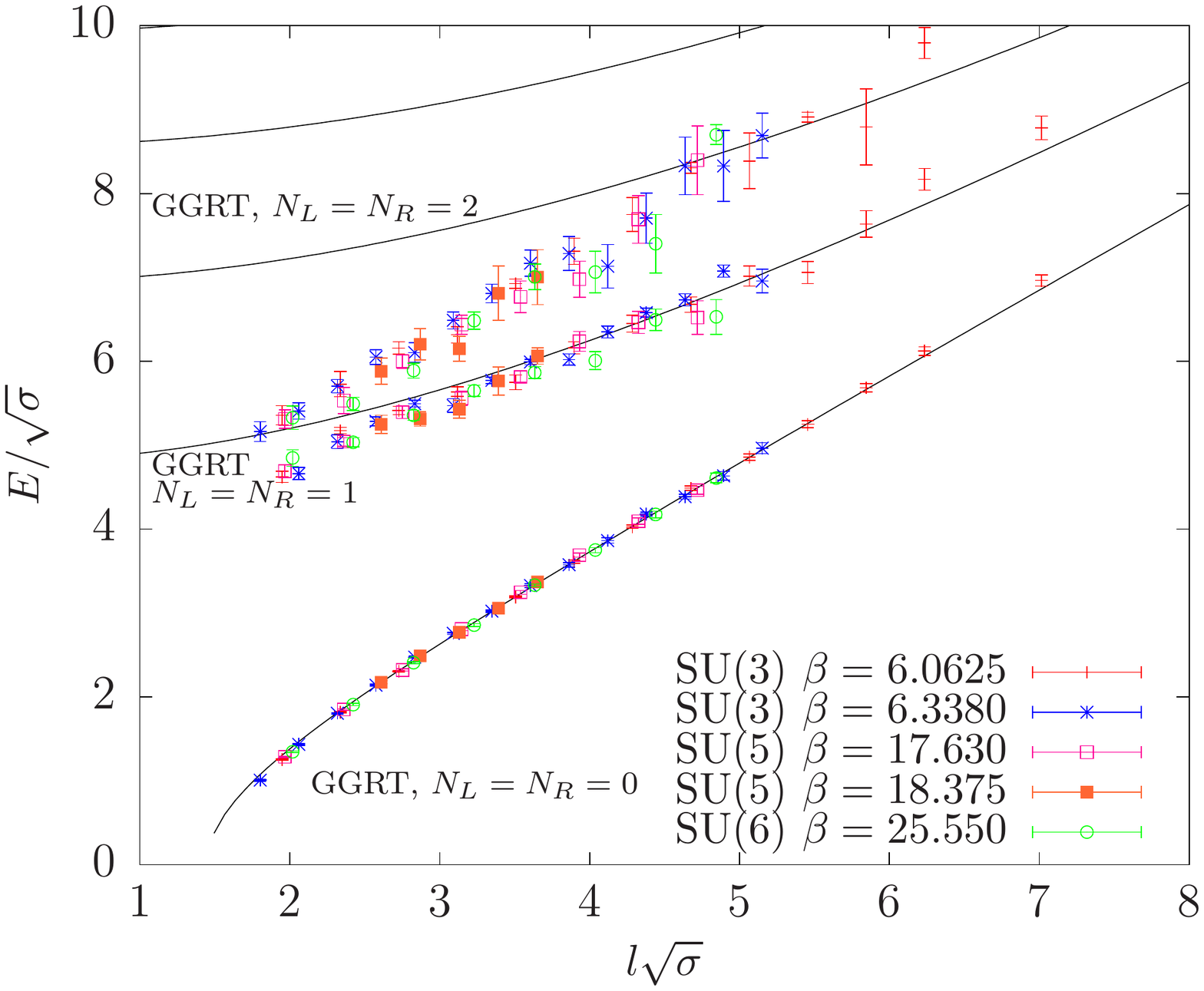} 
    \vspace{-1cm}
    \caption{The ground, first excited and second excited state for $|J_{\rm mod \ 4}|^{P_{\perp}, P_{||}}=0^{++}$ and all the gauge groups used in this work.}
    \label{fig:third_1}
\end{figure}

\begin{figure}[h]
    \centering
    \vspace{-1.5cm}
    \includegraphics[height=10cm]{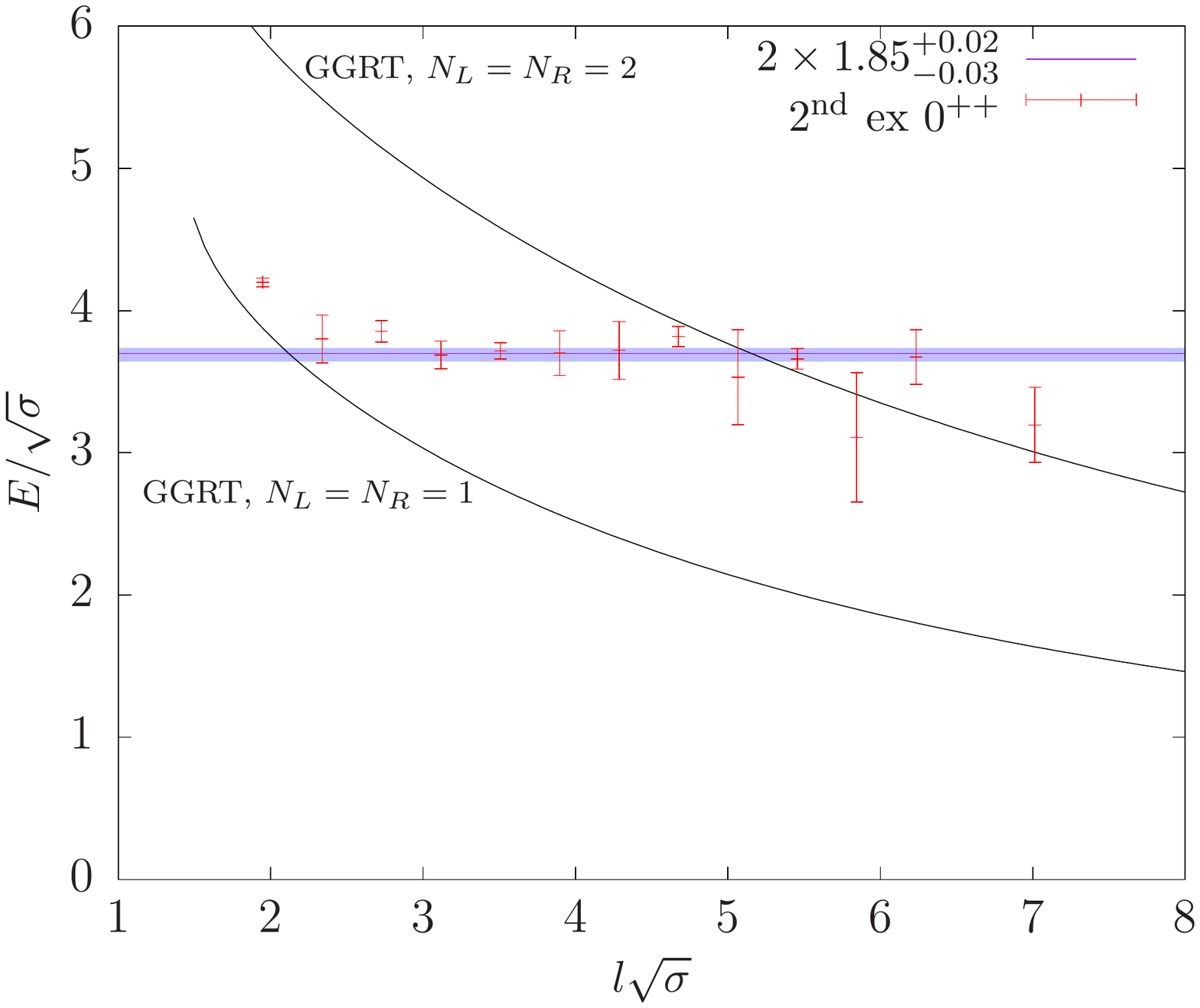} 
    \vspace{-1cm}
    \caption{The second excited state for $|J_{\rm mod \ 4}|^{P_{\perp}, P_{||}}=0^{++}$ with the absolute ground state being subtracted for $SU(3)$ at $\beta = 6.0625$.}
    \label{fig:third_2}
\end{figure}

In Figure~\ref{fig:third_1} we present the second excitation state with quantum numbers $0^{++}$. Above this energy level we get a plethora of states which reflect the multifold degeneracy of the GGRT string for $N_L = N_R = 2$. Strikingly, this state appears to exhibit the same resonance behaviour as the $0^{--}$ ground state: it appears as a constant term coupled to the absolute ground state. This is more obvious if we subtract from this energy level the contribution of the absolute ground state as this appears in Figure~\ref{fig:third_2}. Namely, we observe that this is in agreement with a resonance of mass twice that of the axion. This raises the question whether such a relation is accidental or it has some deeper interpretation. A reasonable expectation would be that this state is a bound state of two axions with a very low binding energy; this scenario is in agreement with the quantum numbers of the state.

\subsubsection{The $q \ne 0$ sector and the appearance of the world-sheet axion}
\label{sec:nonzeromomentum}

\begin{figure}
    \centering
    \vspace{-0.5cm}
    \hspace{-3cm}
    \includegraphics[height=8cm]{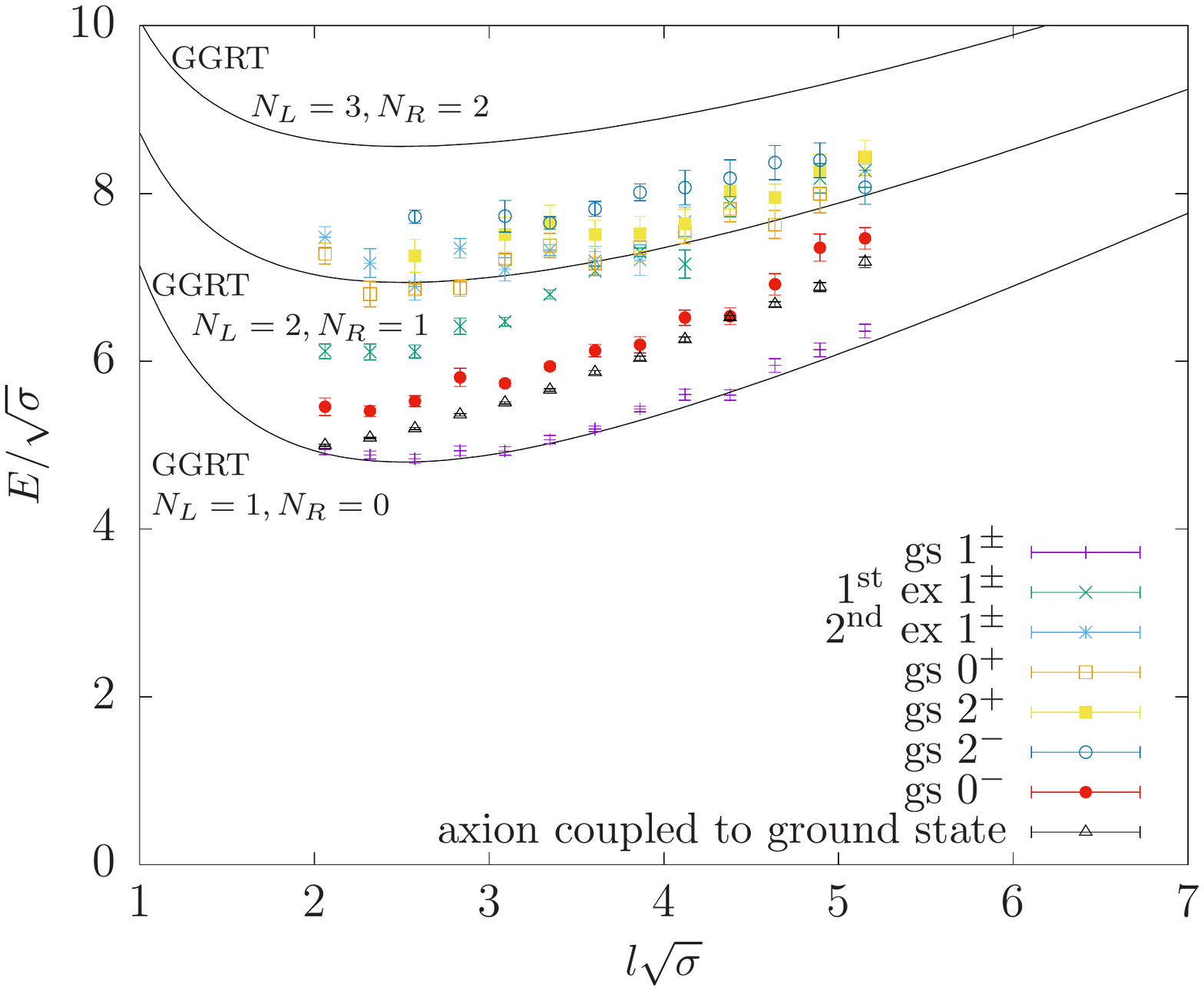} \hspace{-3cm}
    \includegraphics[height=8cm]{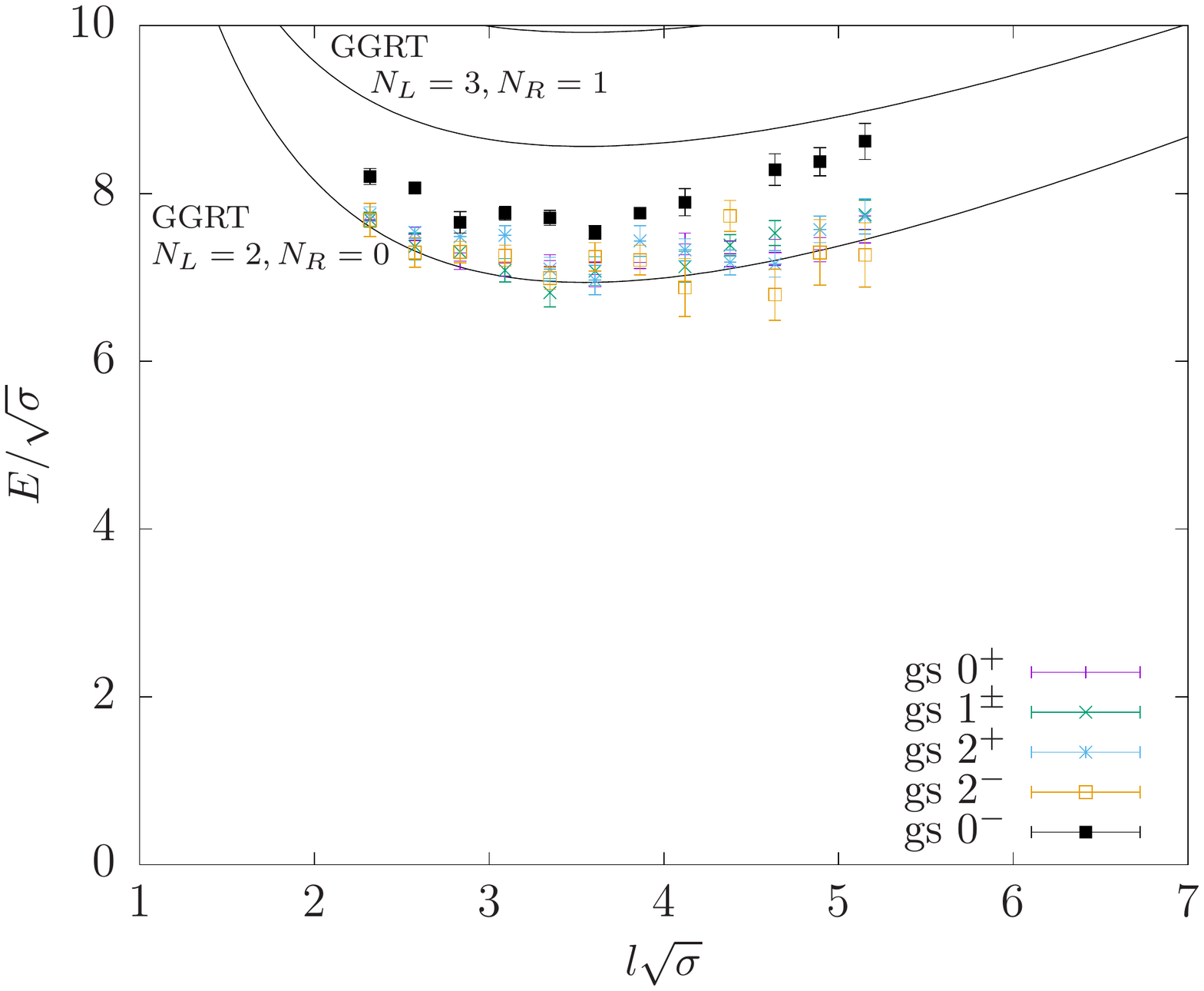} \hspace{-4cm}
    \vspace{-1cm}
    \caption{{\underline{Left Panel}}: The ground, first excited and second excited $1^{\pm}$ states as well as the ground states $0^+$, $0^-$, $2^+$, $2^-$ for a flux-tube with $q=1$ in $SU(3)$, $\beta=6.338$. {\underline{Right Panel}}: The ground states with quantum numbers $0^{+}$, $0^-$, $1^{\pm}$, $2^+$, $2^-$ for a flux-tube with $q=2$ at $SU(3)$, $\beta = 6.338$.}
    \label{fig:fourth}
\end{figure}

In this section we present our results for the $q=1$ and $q=2$ momentum sectors. In the left panel of figure~\ref{fig:fourth} we demonstrate the spectrum  for $q=1$, $SU(3)$ and $\beta=6.338$. Since, the string ground state $N_L=1$, $N_R=0$ can only be created by a single phonon, it has $J=1$. The flux-tube ground state with quantum numbers $1^{\pm }$, $q=1$ appears to be in good agreement with the prediction of the GGRT string. This is in accordance with the results of Ref~\cite{Dubovsky:2013gi}. The next string excitation level, corresponding to $N_L=2$ and $N_R=1$ should be seven-fold degenerate. This should consist of one $0^{+}$, one $0^{-}$, three $1^{\pm}$, one $2^{+}$ and a $2^-$ state. In the left panel of Figure~\ref{fig:fourth} we show the flux-tube ground state with quantum numbers $2^+$, the ground state with $2^-$, the ground state for $0^+$ as well as the first and second excited states with $1^{\pm }$. All the above five states appear to cluster around the GGRT prediction. Furthermore, we demonstrate the ground state for $0^{-}$ which appears to exhibit large deviations from the GGRT string. Since, this state has the same quantum numbers as the pseudoscalar massive excitation the first assumption one could make is that this reflects to the axion. A naive comparison of this state with a relativistic sum of the absolute ground state plus an axion with momentum $2 \pi / l$ is provided in the same figure, demonstrating an approximate agreement with our data for large flux-tubes. This strengthens the scenario of this state being the world-sheet axion.  

In the right panel of Figure~\ref{fig:fourth} we show results for $q=2$, $SU(3)$ and $\beta=6.338$. The string ground state $N_L=2$, $N_R=0$ is expected to be four-fold degenerate. Namely, it is expected to be occupied by states with quantum numbers $0^{+}$, $1^{\pm}$, $2^{+}$ and $2^{-}$. We, thus, extract the flux-tube ground states with the above quantum numbers and observe that they all cluster around the GGRT prediction. The next string excitation level is multi-fold degenerate and should also include a $0^{-}$ state which encodes the quantum numbers of the axion. We extract the flux-tube ground state with quantum numbers $0^-$ and we observe a very similar behaviour as for the case of $q=1$; namely it diverges greatly from the GGRT prediction. 

\subsection{The spectrum of glueballs in the planar limit}
\label{sec:glueballs4D}
At this section of the manuscript we present results for the spectrum of glueballs in $SU(N)$ gauge theories at the continuum limit $a \to 0$ as well as their extrapolations to the $N = \infty$ limit. These spectra have been extracted from calculations on $SU(2)$ and $6$ values of the lattice spacing ($\beta=2.2986, \ 2.3714, \ 2.427, \ 2.509, \ 2.60, \ 2.70$), on $SU(3)$ and $8$ values of the lattice spacing ($\beta=5.6924, \ 5.80, \ 5.8941, \ 5.99, \ 6.0625, \ 6.235, \ 6.3380, \ 6.50$), on $SU(4)$ and $6$ values of the lattice spacing ($\beta=10.70, \ 10.85, \ 11.02, \ 11.20, \ 11.40, \ 11.60$), on $SU(5)$ and $5$ values of the lattice spacing ($\beta=16.98, \ 17.22, \ 17.43, \ 17.63, \ 18.04, \ 18.375$), on $SU(6)$ and $6$ values of the lattice spacing ($\beta=24.67, \ 25.05, \ 25.32, \ 25.55, \ 26.22, \ 26.71$), on $SU(8)$ and $6$ values of the lattice spacing ($\beta=44.10, \ 44.85, \ 45.50, \ 46.10, \ 46.70, \ 47.75$), on $SU(10)$ and $5$ values of the lattice spacing ($\beta=69.20, \ 70.38, \ 71.38, \ 72.40, \ 73.35$) and finally on $SU(12)$ and $5$ values of the lattice spacing ($\beta=99.86, \ 101.55, \ 103.03, \ 104.55, \ 105.95$). As before, critical slowing down introduces systematic errors which should be addressed carefully. Since this refers to technical aspects of the calculation and is, thus, beyond the scope of this presentation, we refer the reader to the actual publication~\cite{Athenodorou:2021qvs}.

For the gauge groups presented above, we calculated glueball masses from the correlators of suitable operators which have been encoded zero momentum $p=0$ by imposing translation invariance. These operators are chosen to have quantum numbers $R^{PC}$ as this has been presented in Section~\ref{sec:lattce_gauge_theory}. 12 different closed loops on the lattice have been used to facilitate the Generalized Eigenvalue Problem; all the loops are presented in Figure~\ref{fig:glueball_operators}. For each different loop all 24 rotations as well as their linear combinations of the traces that transform irreducibly under $R$ have been constructed. By taking the real and imaginary parts of the traces separately,
we build operators with $C=\pm$ respectively. We also calculate the parity inverses of
each of these 12 closed loops, and of their rotations, and by adding and subtracting appropriate
operators from these two sets we form operators for each configuration of the quantum numbers $R$ with $P=\pm$. Once more, states in cubic representations $A_1$ and $A_2$ appear to be one-dimensional, meaning that for each energy level we have only one such state. States in the $E$ representation are doubly degenerate (two-dimensional) and in the $T_1$ and $T_2$ are triply degenerate (three-dimensional).

Lattice simulations are computationally intensive, and for reasons of computational economy we wish to perform calculations on lattice volumes that are small but, at the same time, large enough so that any finite volume effects do not interfere with the physics under investigation. The computational cost of simulating and calculating in $SU(N)$ gauge theories increases approximately as $\propto N^3$ due to the multiplication of two $N \times N$ matrices. Since finite volume corrections are expected to decrease as powers of $1/N$ , we reduce the size in physical units of our lattices as we increase $N$. Due to the technical nature of this topic we refer the reader to our longer manuscript in Ref~\cite{Athenodorou:2021qvs}.

Special attention has been given in ensuring that finite volume effects do not affect the spectrum of the glueballs. There are two types of such finite volume corrections. The first arises when the propagating glueball emits a virtual glueball which propagates around
the spatial torus. The shift caused by the virtual gluballs in the mass of the propagating glueball decreases exponentially in $m_G l_x$ with $l_x$ being the length of the spatial torus. For the glueball calculation we choose $l_x$ so that $am_G \times l_x / a$ is large enough and, thus, 
we can expect this correction to be small. Details on the choice of $l_x$ can be found in Ref.~\cite{Athenodorou:2021qvs}. Similar source of finite volume effects is also present in the confining-string spectrum where in order to ensure that such effects are under control we have chosen the transverse directions of the lattice $l_y=l_z$ to be adequately large.

The second type of finite volume effects in the glueball spectrum includes states composed of several flux-tubes winding around a spatial torus in a singlet state. The lightest of these states will be composed of one winding flux-tube together with a conjugate winding flux-tube and we, thus, refer to it as a `ditorelon'. These states have a non-zero overlap onto the loops (Figure~\ref{fig:glueball_operators}) we use as our glueball operators. Therefore, it can appear as a state in our extracted glueball spectrum. Neglecting interactions between the flux-tubes, the lightest ditorelon will consist of each flux-tube in its ground state with zero momentum and will have an energy, $E_d$, that is twice that of the flux-tube absolute ground state $E_{\rm gs}$, $E_d=2E_{{\rm gs}}$. In principle we expect interactions to shift the energy
but this shift should be small on the volumes we have chosen. Hence, we shall use $E_d\simeq 2E_{{\rm gs}}$ as a rough estimate in searching for these states. The ditorelon ground state contributes only to the $A_1^{++}$ and $E^{++}$ representations. If we allow one or
both of the component flux-tubes to be excited and/or to have non-zero equal and opposite
transverse momenta we can populate other representations and produce towers of states.
However, these excited ditorelon states will be considerably heavier on the lattice volumes
we use. Ditorelon contributions in the $A_1^{++}$ and $E^{++}$ channels have been investigated in detail in the longer write-up. Namely, operators which have been constructed in such a way so that they maximise the overlap onto ditorelon states have been used. This enabled us to identify ditorelon states which appeared in the calculated glueball spectra and ensure that the quoted glueball spectrum consists solely of glueball states.
 
\subsubsection{Continuum masses}
\label{sec:continuum_masses}

For each value of $N$ for  $SU(N)$ we have extracted the low-lying glueball spectra for a range of values of $a(\beta)$. All the masses are expressed in lattice units $aM$, and to transform that to physical units we can take the ratio to the string tension, $a\sqrt{\sigma}$,
that we have simultaneously calculated. We can then extrapolate this ratio to the continuum limit using the standard Symanzik effective action analysis 
that tells us that for our lattice action the leading correction at tree-level is $O(a^2)$:
\begin{eqnarray}
\frac{aM(a)}{a\sqrt{\sigma(a)}} = \frac{M(a)}{\sqrt{\sigma(a)}}
=
\frac{M(0)}{\sqrt{\sigma(0)}} + a^2\sigma(a) + O(a^4).
\label{eqn_MKcont}
\end{eqnarray}
In the above expression we have used the calculated string tension, $a^2\sigma(a)$, as the $O(a^2)$ correction. Clearly, we could use any other calculated energy, and this would differ at $O(a^4)$ in Equation~(\ref{eqn_MKcont}). We choose to use $a^2\sigma(a)$ since we can extract it with small errors.

\begin{figure}
    \centering
    \includegraphics[height=5.9cm]{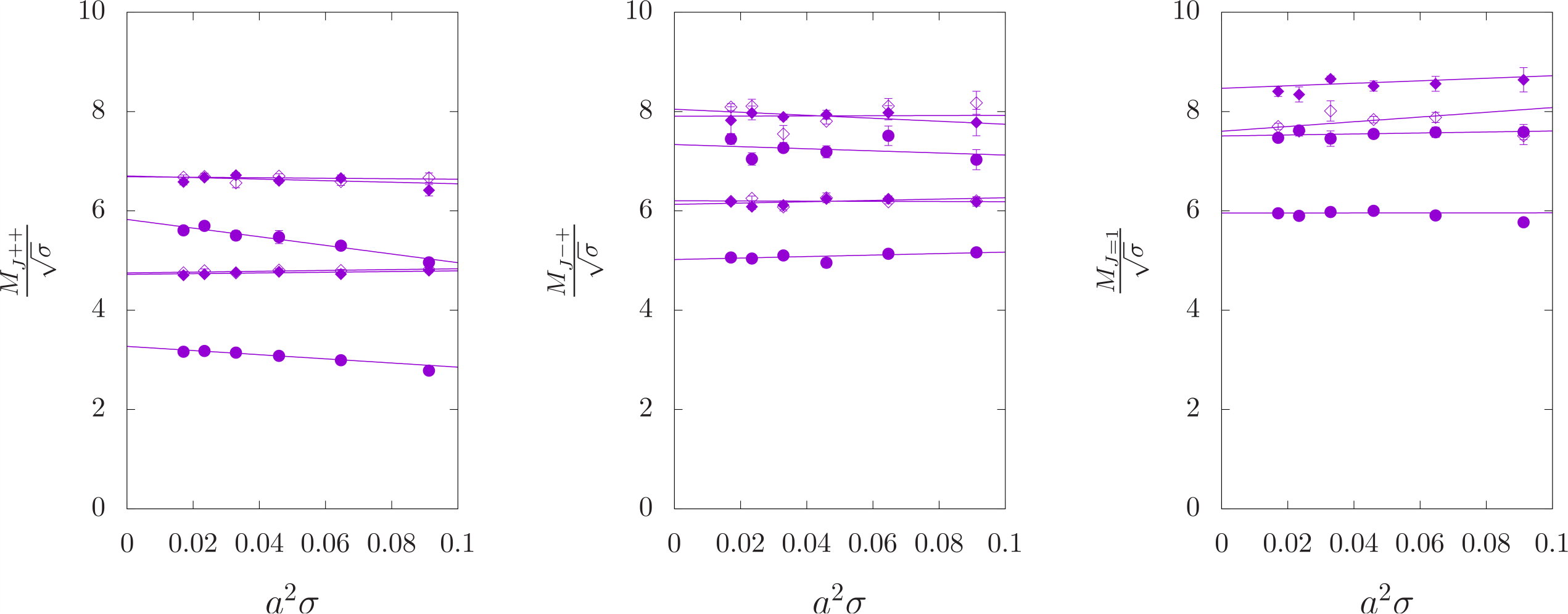} 
    \caption{\label{fig:continuum_masses} \underline{Left Panel:} Lightest two glueball masses in the $A_1^{++}$ ($\bullet$), $E^{++}$ ($\blacklozenge$)
  and $T_2^{++}$ ($\lozenge$) sectors, in units of the string tension. Lines are linear
  extrapolations to the continuum limit. In that limit the  $A_1^{++}$ states become the
  lightest two $J^{PC}=0^{++}$ scalar glueballs while the doublet $E^{++}$ and triplet $T_2^{++}$
  pair up to give the five components of each of the lightest two $J^{PC}=2^{++}$ glueballs. \underline{Middle Panel:} Lightest two glueball masses in the $A_1^{-+}$ ($\bullet$), $E^{-+}$ ($\blacklozenge$)
  and $T_2^{-+}$ ($\lozenge$) sectors, in units of the string tension. Lines are linear
  extrapolations to the continuum limit. In that limit the  $A_1^{-+}$ states become the
  lightest two $J^{PC}=0^{-+}$ pseudoscalar glueballs while the doublet $E^{--+}$ and triplet $T_2^{-+}$
  pair up to give the five components of each of the lightest two $J^{PC}=2^{-+}$ glueballs.  \underline{Right Panel:} Lightest two glueball masses in the $T_1^{+-}$ ($\bullet$) representation
  and the lightest ones in the $T_1^{-+}$ ($\blacklozenge$)
  and $T_1^{--}$ ($\lozenge$) representations, in units of the string tension. Lines are linear
  extrapolations to the continuum limit. In that limit the  $T_1^{+-}$ states become the
  lightest two $J^{PC}=1^{+-}$ glueballs while the other two becomes the
  $1^{-+}$ and $1^{--}$ ground state glueballs. All three plots for $SU(4)$. }
\end{figure}

In the left panel of Figure~\ref{fig:continuum_masses} we demonstrate our extrapolations of the lightest two $A_1^{++}$,  $E^{++}$ and $T_2^{++}$ states for $SU(4)$. These states are of particular importance because, as explained in Section~\ref{sec:quantum_numbers_glueballs}, they correspond to the lightest two $J^{PC} = 0^{++}$ and $2^{++}$ states. As can be seen all the fits appear to be linear, confirming the expression provided in Equation~\ref{eqn_MKcont}. In the middle panel of Figure~\ref{fig:continuum_masses} we show the corresponding plot for $P= -$ which corresponds to the lightest two $J^{PC} = 0^{-+}$ and $2^{-+}$ states. The lightest states have very plausible continuum extrapolations, although the excited states, which are heavier than those for $P=+$, begin to show a large scatter character indicating a poor fit.
In the right panel of Figure~\ref{fig:continuum_masses} we present the continuum extrapolations of various $T_1^{PC}$ states that correspond to $J=1$, and again we observe that the fits appear to be convincing for the lighter states and quite plausible for the heavier states. Clearly, we can safely state that most of the states exhibit small lattice artifacts since the slopes of the continuum extrapolations appear to be small.

Finally, in the left column of Figure~\ref{fig:glueball_masses} we provide the extrapolated results for, the phenomenologically most interesting case of  $SU(3)$ and all the different irreducible representations configured  by representations $A_1$, $A_2$, $E$, $T_1$ and $T_2$ as well as by $P=\pm$ and $C=\pm$.

\subsubsection{Large-$N$ extrapolations}
\label{sec:large-N-extrapolations}
\begin{figure}
    \centering
    \vspace{-2cm}
    \rotatebox{0}{
    {\includegraphics[height=5.8cm]{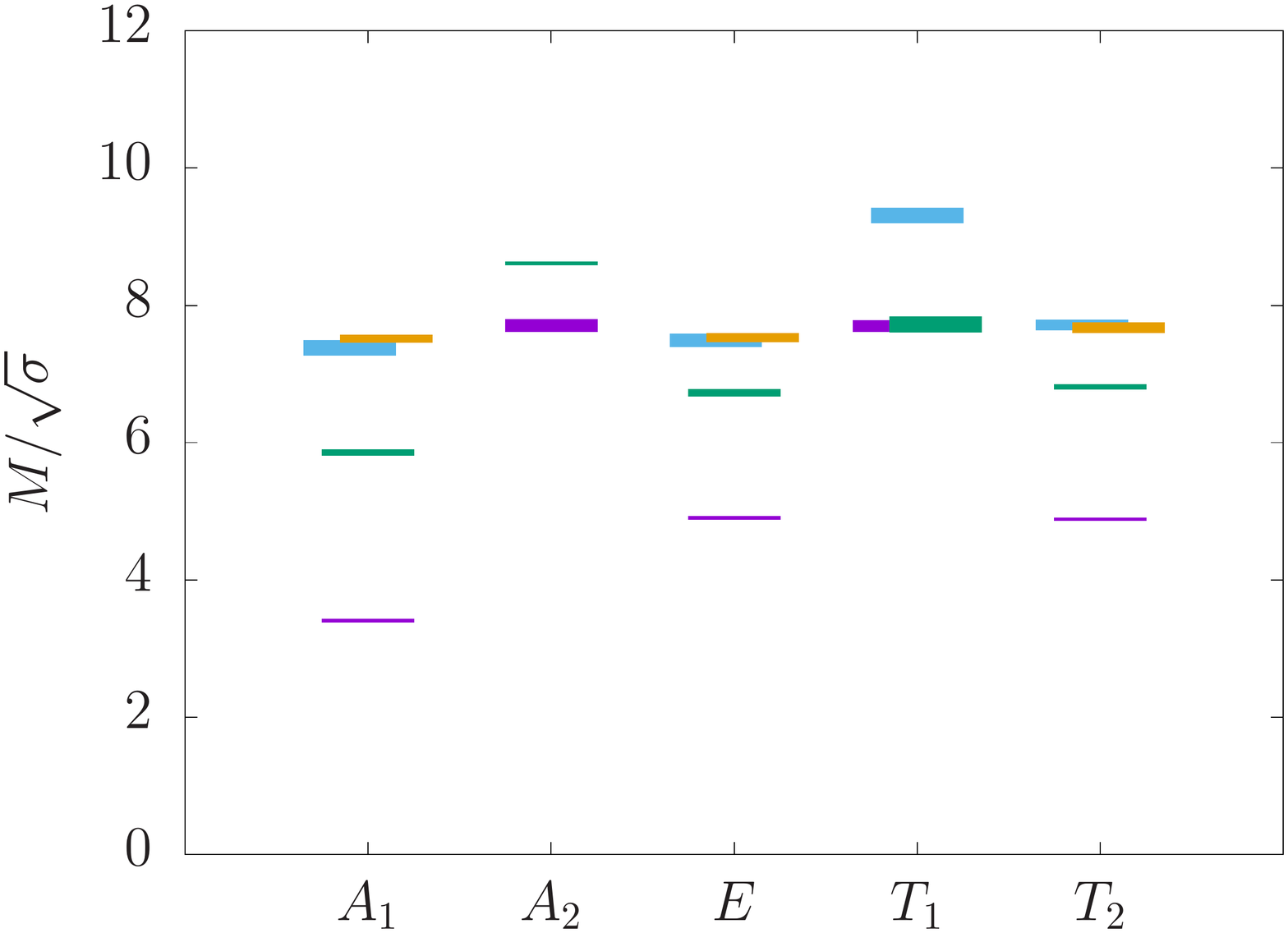}} \includegraphics[height=5.8cm]{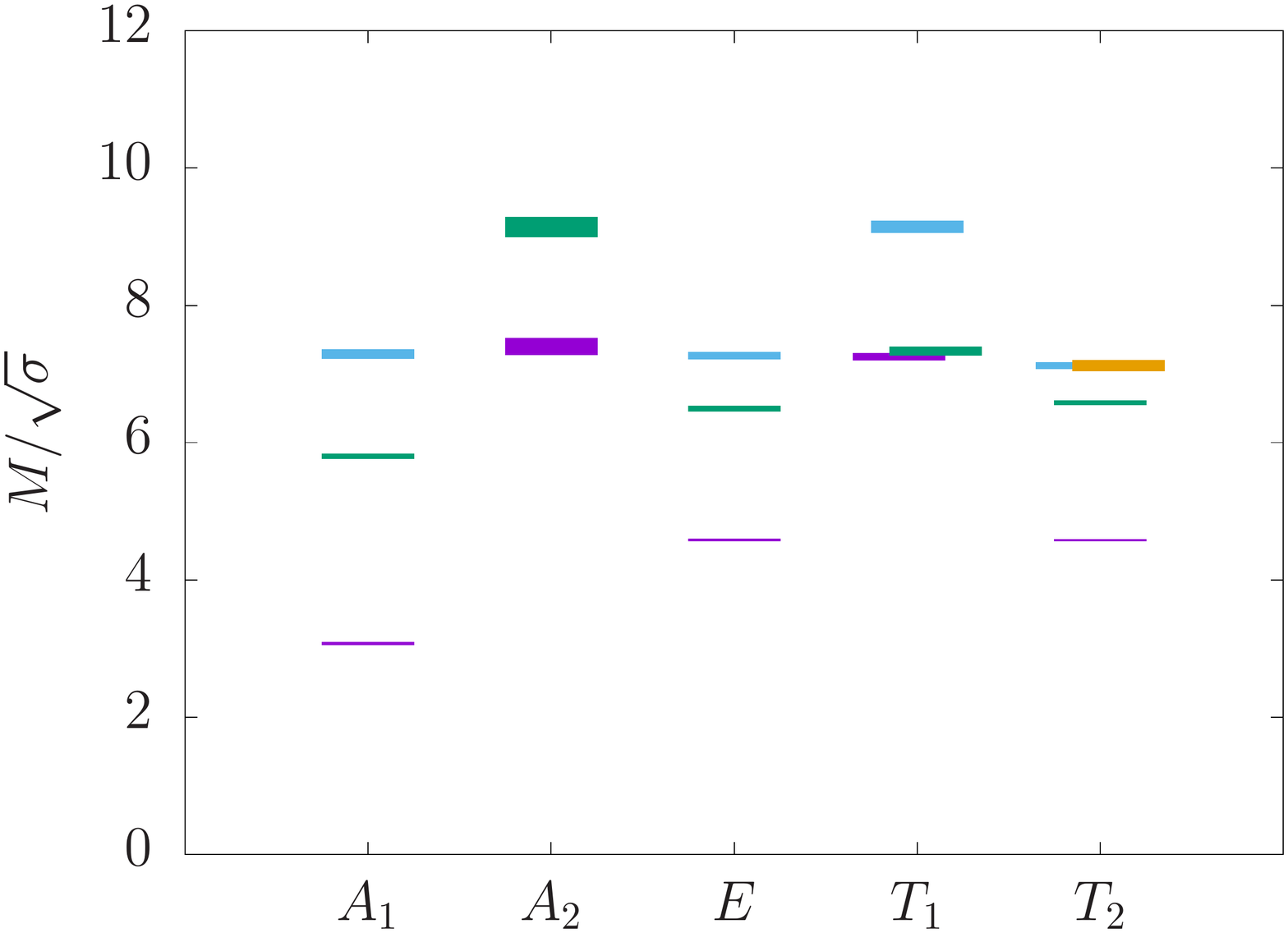} \put(-335,160){$SU(3)$}\put(-125,160){$SU(\infty)$}\put(-310,35){$P=+, C=+$}\put(-95,35){$P=+, C=+$}  }
    \vspace{-1.5cm}
    
    \rotatebox{0}{
    \includegraphics[height=5.8cm]{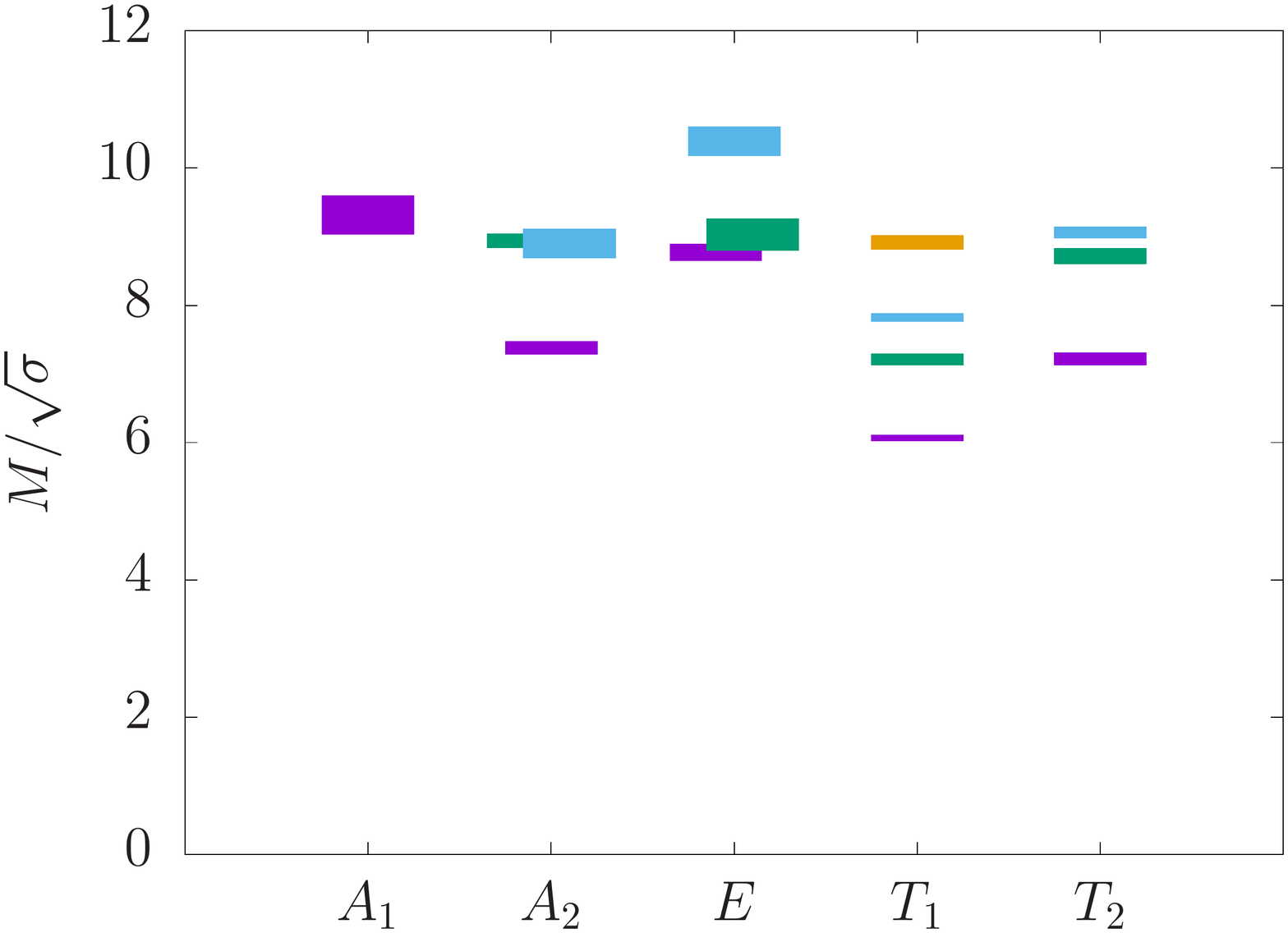} \includegraphics[height=5.8cm]{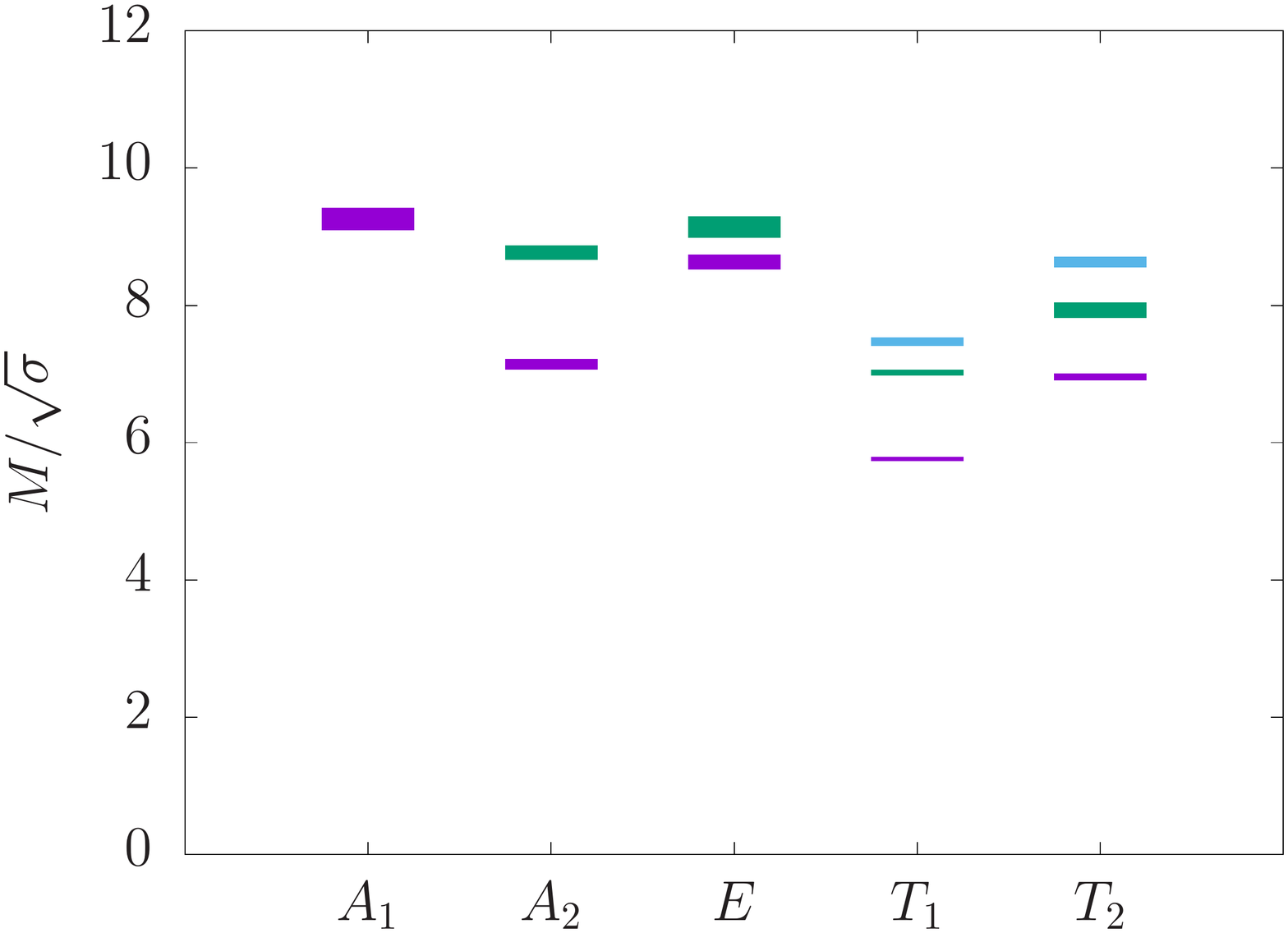} 
    \put(-310,35){$P=+, C=-$}\put(-95,35){$P=+, C=-$}  }
    \vspace{-1.5cm}
    
    \rotatebox{0}{
    \includegraphics[height=5.8cm]{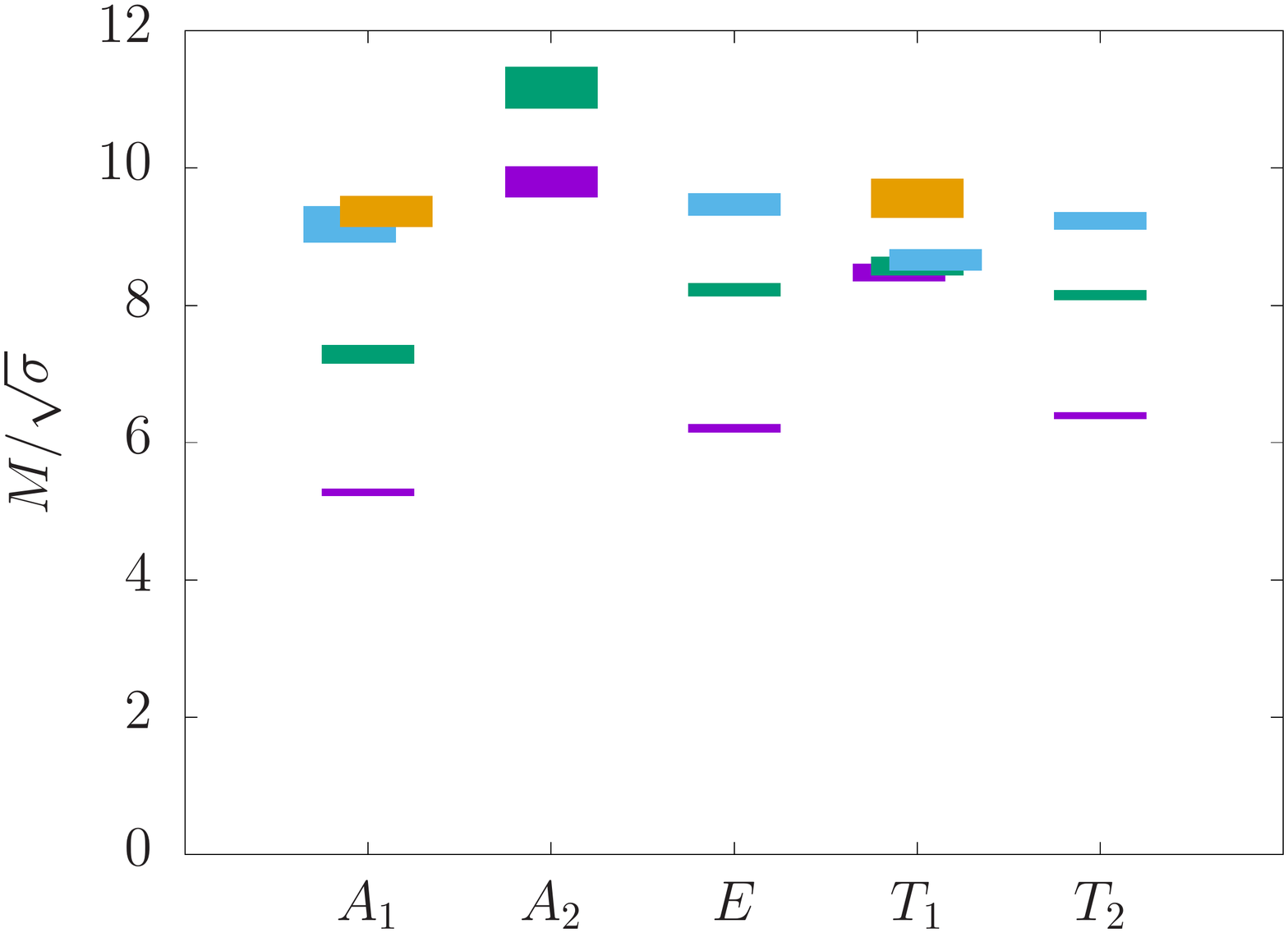} \includegraphics[height=5.8cm]{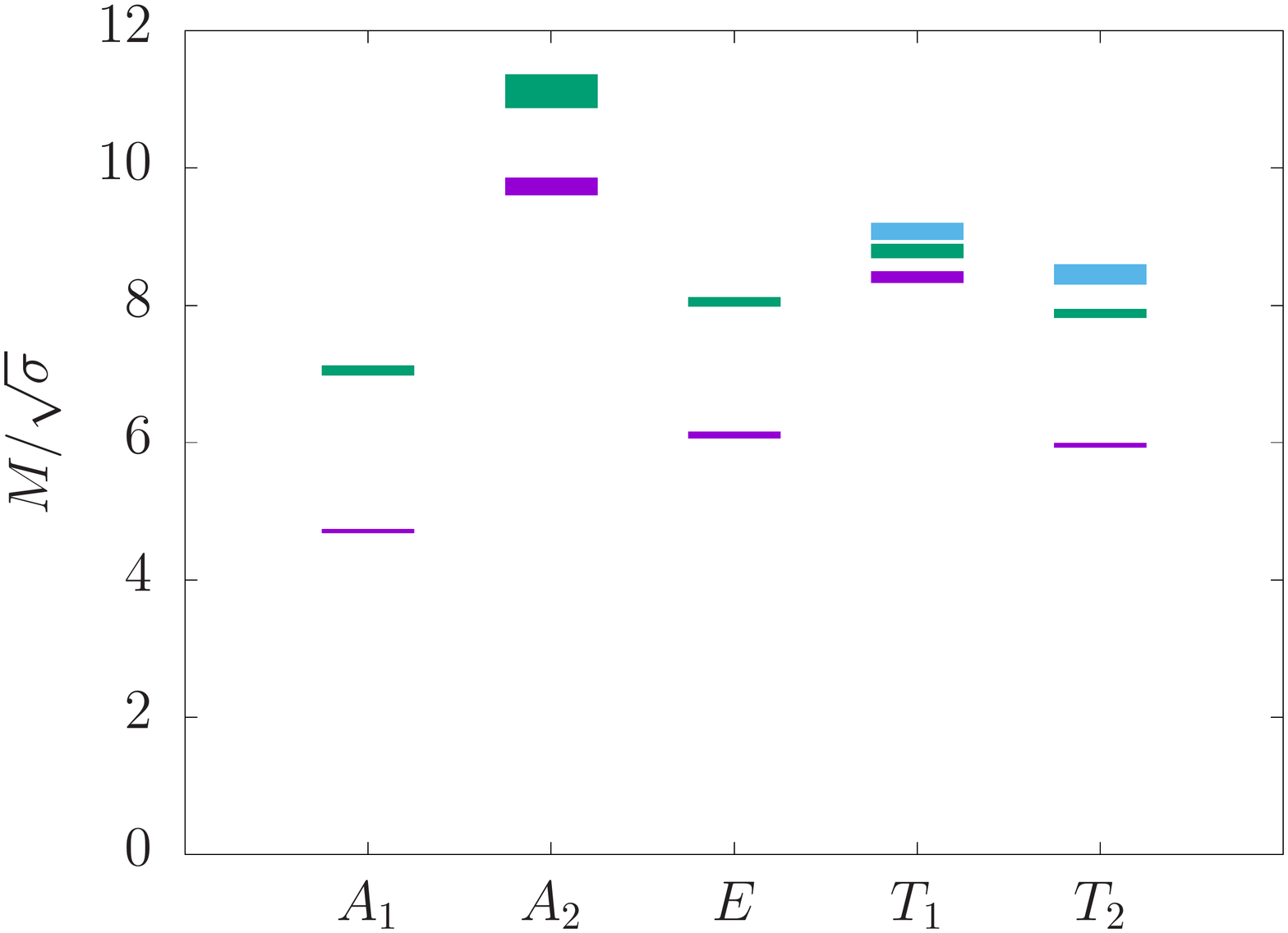} 
    \put(-310,35){$P=-, C=+$}\put(-95,35){$P=-, C=+$}  }
     \vspace{-1.5cm}
    
    \rotatebox{0}{
    \includegraphics[height=5.8cm]{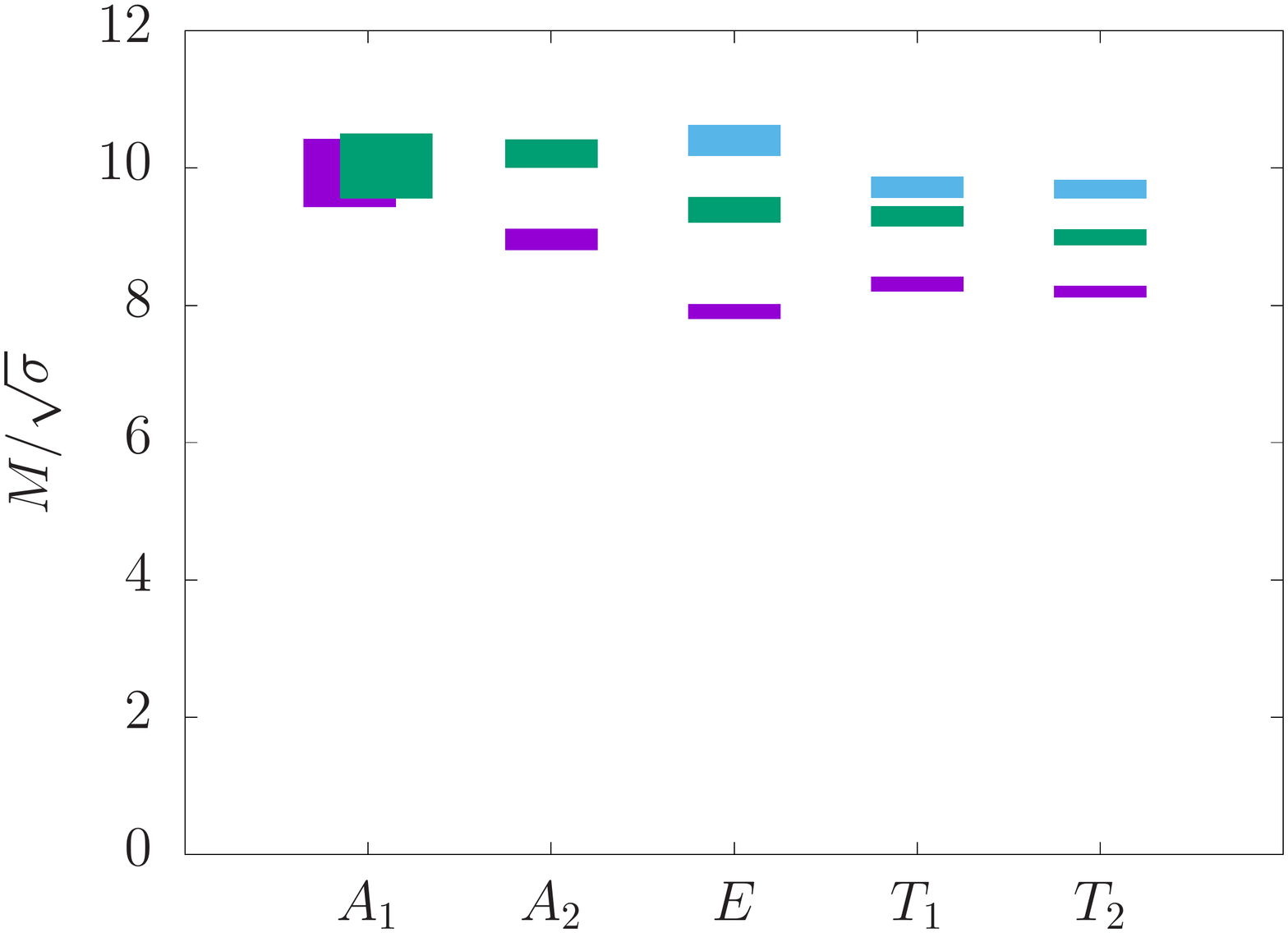} \includegraphics[height=5.8cm]{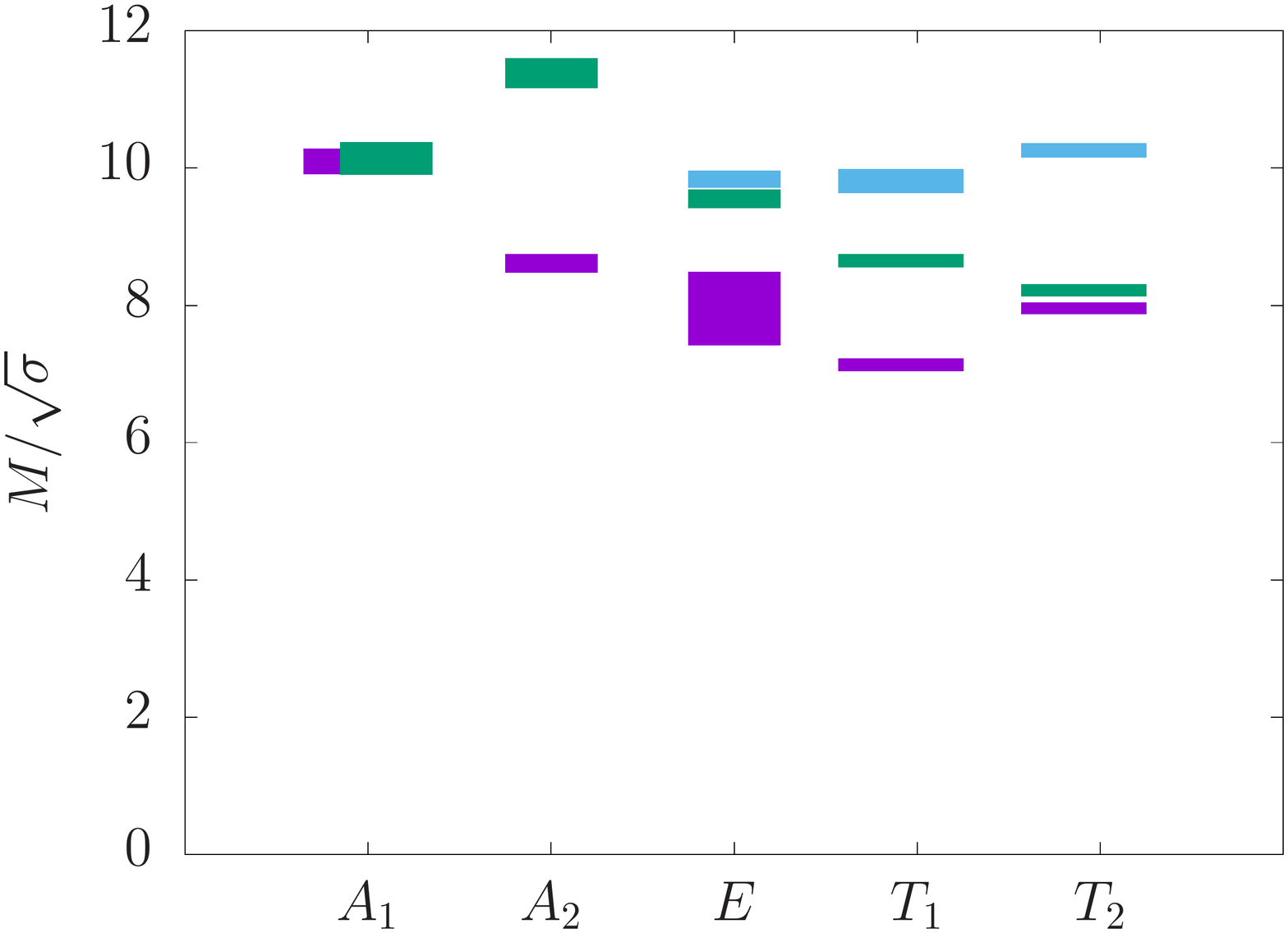}
    \put(-310,35){$P=-, C=-$}\put(-95,35){$P=-, C=-$}  }
    \caption{Glueball Masses for $SU(3)$ (left panel) and $SU(\infty)$ (right panel) for the five irreducible representations of the cubic group of rotations as well as for the configuration of quantum numbers $P, C$.}
    \label{fig:glueball_masses}
\end{figure}

Undoubtedly, from a phenomenological point of view, the most interesting calculation of glueball spectra is that for $SU(3)$ presented in the left panel of Figure~\ref{fig:glueball_masses}. Thus, a whole paper to that case~\cite{Athenodorou:2020ani} has been devoted. However, from a theoretical point of view, the most interesting glueball spectra
are those of the $SU(N\to\infty)$ theory since the theoretical simplifications
in that limit make it the most likely case to be accessible to analytic
solution, whether complete or partial.

To extract the $N=\infty$ spectrum from the data obtained for the sequence of values of $N$, one can use the fact that in the pure gauge theory, as explained in Section~\ref{sec:largeN}, the leading correction is $O(1/N^2)$. So we can extrapolate the continuum mass ratios using the formula
\begin{eqnarray}
\left.\frac{M_i}{\sqrt{\sigma}}\right|_{N}
=
\left.\frac{M_i}{\sqrt{\sigma}}\right|_{\infty}
+ \frac{c_i}{N^2} + O\left(\frac{1}{N^4}\right).
\label{eqn_MKN}
\end{eqnarray}
The results of the extrapolation to the $N \to \infty$ are presented in the right column of Figure~\ref{fig:glueball_masses} as well as in Table~\ref{tab:table_MK_R_SUN}. 

Most of the fits are for $N \geq 2$ or for $N \geq 3$ but some fits are over a more restricted range of $N$ mainly for technical reasons. For instance, the $A_2^{++}$ ground state, has been fitted to $N \geq 4$, the $T_2^{-+}$ second excited state, has been fitted to $N \geq 4$, the $T_2^{--}$ ground state, has been fitted to $N \geq 4$, and finally the $A_2^{+-}$ ground has been fitted to $N \leq 8$.

\begin{figure}
    \centering
    \includegraphics[height=5.9cm]{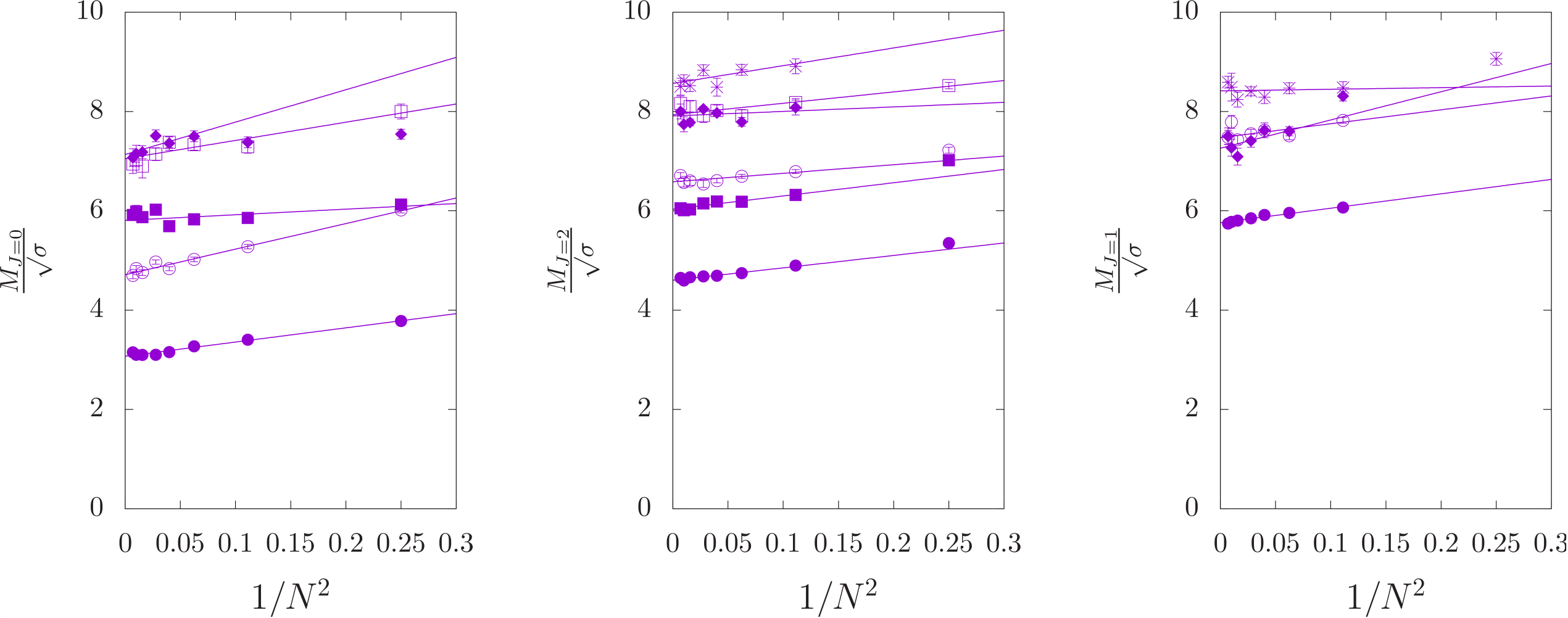} 
    \caption{\label{fig:large_N} \underline{Left Panel:} Continuum masses of the lightest ($\bullet$) and first excited ($\blacksquare$) $J^{PC}=0^{++}$ scalars and of the lightest ($\circ$) and first excited ($\square$)
  $0^{-+}$ pseudoscalars, in units of the string tension. The state denoted by
  $\blacklozenge$ is either the $4^{++}$ ground state or the second excited $0^{++}$.
  With extrapolations from values in the range $N\in[2,12]$ to $N=\infty$. 
  \underline{Middle Panel:} Continuum masses of the lightest ($\bullet$) and first excited ($\circ$)
  $J^{PC}=2^{++}$ tensors, the lightest ($\blacksquare$) and first excited ($\square$)
  $2^{-+}$ pseudotensors, the lightest $2^{+-}$ ($\ast$), and the lightest
  $2^{--}$ ($\blacklozenge$), all in units of the string tension.
  With extrapolations to $N=\infty$ from $N\leq 12$.
  \underline{Right Panel:} Continuum masses of the lightest ($\bullet$) and first excited ($\circ$)
  $J^{PC}=2^{++}$ tensors, the lightest ($\blacksquare$) and first excited ($\square$)
  $2^{-+}$ pseudotensors, the lightest $2^{+-}$ ($\ast$), and the lightest
  $2^{--}$ ($\blacklozenge$), all in units of the string tension.
  With extrapolations to $N=\infty$ from $N\leq 12$. }
\end{figure}

From the practical point of view the most important extrapolations are for those states to which we are able to assign a continuum spin. The extrapolation of these states are presented in the three panels of Figure~\ref{fig:large_N} for states with $J=0,2,1$ respectively. Furthermore, the extrapolated  corresponding glueball masses are given in Table~\ref{table_MK_J_SUN}.

Judging by the behaviour of the extrapolations, the mass ratios appear to be described to an adequate level by Equation~\ref{eqn_MKN} with slopes which are relatively small. This suggests that, glueball spectrum in $SU(3)$ can be approximated to a good extent by the spectrum of $SU(\infty)$. 

\begin{table}[htb]
\centering
\begin{tabular}{c|c|c|c|c} \hline \hline
  $R$   & $P=+,C=+$ & $P=-,C=+$ &  $P=+,C=-$   &  $P=-,C=-$   \\ \hline \hline
$A_1$  & 3.072(14)         & 4.711(25)            & 9.26(16)        &  10.10(18) \\
    & 5.805(31) & 7.050(68)             &                &  10.14(23) \\
    & 7.294(63)         &                       &                &   \\ \hline
$A_2$  & 7.40(12) & 9.73(12)              & 7.142(75)  & 8.61(13) \\
    & 9.14(14)   & 11.12(24)             & 8.77(10)      &  11.38(21)   \\ \hline
$E$   & 4.582(14)         & 6.108(44)             & 8.63(10)      &  7.951(53) \\
    & 6.494(33)  & 8.051(60)             & 9.14(15)      &  9.55(13)                   \\
    & 7.266(50)  &                       &               &  9.84(12)                 \\ \hline
$T_1$  & 7.250(47)         & 8.412(76)             & 5.760(25)     & 7.134(86)  \\
    & 7.337(60)  & 8.79(10)              & 7.020(39)     & 8.65(9)  \\
    & 9.142(82)         & 9.08(12) & 7.470(55)     & 9.81(17)  \\
    &                   &                       & 8.422(84)     &             \\ \hline
$T_2$  & 4.578(11)         & 5.965(28)             & 6.957(41)     &  7.96(8) \\
    & 6.579(30)         & 7.883(57)             & 7.93(11)      &  8.22(8)            \\
    & 7.121(45)         & 8.45(14)     & 8.63(7) & 10.26(10)  \\
    & 7.122(76)         &     &     &  \\ \hline \hline
\end{tabular}
\caption{Continuum glueball masses in units of the string tension,
  in the limit $N\to\infty$. Labels are $R$ for the representations of the rotation symmetry of a cube,
  $P$ for parity and  $C$ for charge conjugation.}
\label{tab:table_MK_R_SUN}
\end{table}

\begin{table}[htb]
\centering
\begin{tabular}{c|c|c|c|c} \hline  \hline
  $J$      & $P=+,C=+$   & $P=-,C=+$   &  $P=+,C=-$   &  $P=-,C=-$   \\ \hline \hline
 $0$ {gs}  & 3.072(14) & 4.711(26) &  $\ge 9.26(16)$     & $\ge 10.10(18)$    \\
 $0$ {ex}  & 5.845(50) & 7.050(68) &                     &    \\
 $2$ {gs}  & 4.599(14) & 6.031(38) & 8.566(76)           &  7.910(56)  \\
 $2$ {ex}  & 6.582(36) & 7.936(54) &  $\ge 9.14(15)$     & $\ge 9.55(13)$    \\
 $1$ {gs}  & $\ge 9.14(9)$ & 8.415(76)  & 5.760(25)      & 7.26(11)   \\
 $1$ {ex}  &           &                & 7.473(57)      & $\ge 8.65(9) $  \\
 $3$ {gs}  & 7.263(56) & $\ge 9.73(12)$ & 6.988(41)      & $\ge 8.61(13) $   \\
 $4$ {gs}  & 7.182(71) & $\ge 8.79(10)$ & $\ge 9.26(16)$ & $\ge 10.10(18)$   \\  \hline \hline
\end{tabular}
\caption{Large-$N$ extrapolation of continuum glueball masses, in units of the string tension,
  for those configurations of $J^{PC}$ we can identify, with lower bounds in those
  cases where this is not possible. Ground state denoted by gs,
  first excited state by ex.}
\label{table_MK_J_SUN}
\end{table}

\section{Conclusions}
\label{sec:conclude}

In this work we have improved extensively the extraction of and, thus, our knowledge on the spectrum of the closed confining string. The majority of the states appearing in the spectrum are string-like, in the sense they can be adequately approximated by a low energy effective string theory. In addition a small sector of the excitation spectrum appears to be massive resonances which can be interpreted as an axion on the world-sheet of the theory. We concluded to the above by the resonance character of the $0^{--}$, $q=0$ ground state which appears to be an axion coupled to the string's absolute ground state, by the $0^{++}$ second excited state which can be interpreted as a bound state of two axions with a very low binding energy coupled to the absolute ground state as well as by the $0^-$ $q=1,2$ ground states which also have an axion character. Furthermore, states with axionic character can also be identified in other irreducible representations such as $|J_{\rm mod \ 4}|^{P_{\perp}} = 1^{\pm}$; this is a matter of presentation in the longer write-up~\cite{Athenodorou:2021vkw}. Finally, the spectrum of the confining string, including the axionic modes appears to have insignificant large-$N$ effects.

Additionally, this manuscript presents the calculation of the glueball spectra of a range of $SU(N)$ gauge theories, in the continuum limit as well as their extrapolations to the theoretically interesting $N \to \infty$ limit. This investigation provided the first calculation of the masses of the ground states in all the $R^{PC}$ irreducible representations, as well as some excited states in most such channels, in the continuum limit of the $SU(\infty)$. These results have improved existing calculations~\cite{variational,Lucini:2001ej,Lucini:2010nv} while largely confirming existing results. The main conclusion of this work is that $SU(3)$ is close enough to $SU(\infty)$. Namely, the $J^{PC}=0^{++}$ scalar ground state has a mass of $M_{0^{++}}\sim 3.41 \sqrt{\sigma}$ for $SU(3)$ to $M_{0^{++}}\sim 3.07 \sqrt{\sigma}$ for $SU(\infty)$, the next heavier glueballs are the tensor with a mass of $M_{2^{++}} \sim 1.5 M_{0^{++}}$ and the pseudoscalar $0^{-+}$ which appears to be nearly degenerate with the tensor. Moving higher in energies, we encounter
the $1^{+-}$ with $M_{1^{+-}} \sim 1.85 M_{0^{++}}$, and this is the only relatively light $C=-$ state. With approximately the same mass comes the first excited $0^{++}$ state and then the lightest pseudotensor with $M_{2^{-+}} \sim 1.95 M_{0^{++}}$ follows.  All other states are heavier than twice the lightest scalar, with most of the $C = -$ ground states being very much heavier.

\section*{Acknowledgments}
AA is indebted to the organisers, especially to Dimitrios Giataganas for the invitation to deliver a talk. Furthermore, AA would like to express his gratitude to Michael Teper, with whom this work has been carried out, for the rewarding collaboration as well as for providing a critical reading of the manuscript. Moreover, AA thanks S.~Zafeiropoulos for sharing feedback on the manuscript. AA would also like to thank S.~Dubovsky, V.~Gorbenko, J.~Sonnenschein, E.~Kiritsis and K.~Hashimoto for interesting discussions. AA has been financially supported by the European Union's Horizon 2020 research and innovation programme ``Tips in SCQFT'' under the Marie Sk\l odowska-Curie grant agreement No. 791122 as well as by the Horizon 2020 European research infrastructures programme "NI4OS-Europe" with grant agreement no. 857645.  Numerical simulations have been carried out in the Oxford Theoretical Physics cluster.

\end{document}